\RequirePackage{fix-cm}
\documentclass[twocolumn]{svjour3}          
\smartqed  
\usepackage{graphicx}
\usepackage{xcolor}
\usepackage{color}
\usepackage[caption=false]{subfig}

\usepackage[switch]{lineno}

\begin{document}

\title{A Generalized Approach to Longitudinal Momentum Determination in Cylindrical Straw Tube Detectors
}


\author{Walter Ikegami Andersson \and
        Adeel Akram \and
        Tord Johansson \and
        Ralf Kliemt \and
        Michael Papenbrock \and
        Jenny Regina \and
        Karin Sch\"onning \and
        Tobias Stockmanns
}



\institute{
        Walter Ikegami Andersson \and Adeel Akram \and Tord Johansson \and Michael Papenbrock \and Jenny Regina \and Karin Sch\"onning \at Uppsala Universitet, Institutionen för fysik och astronomi,{ \bf Uppsala}, Sweden \and
        Ralf Kliemt \at Helmholtz-Institut Mainz,{ \bf Mainz}, Germany  \and
        Tobias Stockmanns \at Forschungszentrum Jülich, Institut für Kernphysik,{ \bf Jülich}, Germany
}

\date{Received: date / Accepted: date}

\maketitle

\begin{abstract}
The upcoming PANDA experiment at FAIR will be among a new generation of particle physics experiments to employ a novel event filtering system realised purely in software, \textit{i.e.} a software trigger. To educate its triggering decisions, online reconstruction algorithms need to offer outstanding performance in terms of efficiency and track quality. We present a method to reconstruct longitudinal track parameters in PANDA's Straw Tube Tracker, which is general enough to be easily added to other track finding algorithms that focus on transversal reconstruction. For the pattern recognition part of this method, three approaches are employed and compared: A combinatorial path finding approach, a Hough transformation, and a recursive annealing fit. In a systematic comparison, the recursive annealing fit was found to outperform the other approaches in every category of quality parameters and reaches a reconstruction efficacy of 95\% and higher.
\end{abstract}

\section{Introduction}
\label{sec:introduction}

Modern experimental particle physics would be unthinkable without the aid of computers.
Early detectors would yield visual event signatures that could be discerned by the human eye (\textit{e.g.} \cite{cloud}). Later, algorithms were introduced for the identification of particle trajectories, not the least due to the ever-increasing amounts of data.
With time, both accelerator and detector technology improved, leading to larger event and data rates as well as more complex signatures. As a result, the explored physics landscape grew larger and a considerable part of the "low-hanging fruits" could be collected. We have now reached a point where most of the interesting physics signals are not only rare, but also look so similar to the background that they can be difficult to distinguish. Intelligent software that can find the proverbial needle in the all-growing haystack, is a prerequisite for further progress in particle physics.

In this article, we contribute with one piece of the puzzle by presenting a method for reconstructing longitudinal track components in a cylindrical straw tube detector located in a solenoid magnetic field. While many existing methods are adapted for tracks originating from the beam-target interaction point, this method applies to all tracks, also those from the decays of long-lived particles such as hyperons. We implement this method for the PANDA experiment, a next-generation facility for hadron physics that is currently under construction at FAIR in Darmstadt, Germany. We put particular emphasis on resolving the spatial ambiguity that arises during track reconstruction in a cylindrical drift detector. Three different algorithms have been tested for this purpose.

\section{The PANDA detector}
\label{sec:the_panda_detector}

\begin{figure*}
    \centering
    \includegraphics[width=\linewidth]{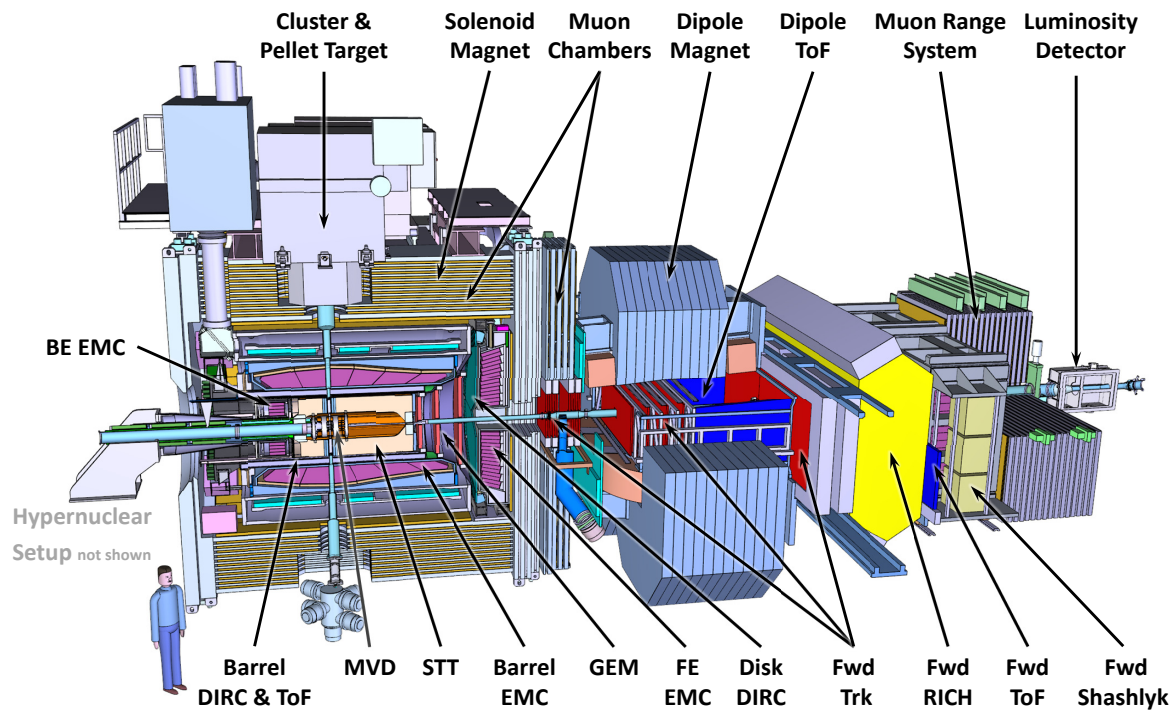}
    \caption{The PANDA detector.}
    \label{fig:panda_full}
\end{figure*}

PANDA (antiProton ANnihilations at DArmstadt)
, shown in Fig.~\ref{fig:panda_full}, is designed as a fixed-target, multi-purpose detector with a wide variety of physics channels and different reaction topologies in mind. 
To enable the exclusive measurement, \textit{i.e.} reconstruction of all final state particles emerging from a given reaction, PANDA will cover almost 4$\pi$ of the solid angle and comprise a range of sub-detectors for the reconstruction of both charged and neutral particles. The subdetectors are grouped into two main segments: The target spectrometer, employing a solenoid magnetic field and covering the polar angle range $10^{\circ} < \theta < 160^{\circ}$, and the forward spectrometer, using a dipole magnetic field and covering the region of $\theta < 10^{\circ}$ and $\theta < 5^{\circ}$ in horizontal and vertical direction, respectively. The purpose of the method presented in this work is to reconstruct the longitudinal track component of charged particles traversing the target spectrometer. 
It utilises the Micro Vertex Detector (MVD) as well as the Straw Tube Tracker (STT).

\subsection{The Micro Vertex Detector}
\label{sec:micro_vertex_detector}


The MVD is a cylindrical pixel- and double-sided strip detector enclosing the interaction point. Its main purpose is to resolve the location of secondary decays of long-lived particles such as hyperons or hadrons with open-charm. 

The MVD consists of a barrel and a forward part. The barrel part is made of four cylindrical detector layers; the innermost layer has a radius of 2.5 cm and the outermost a radius of 13.5 cm. The barrel covers a scattering angle range between $40^{\circ}$ and $150^{\circ}$. The two inner layers consist of silicon hybrid pixels of dimension $100 \times 100 \, \mu$m$^2$. The two outer layers consist of double sided strip detectors. The forward part is made of six disk layers and covers a scattering angle range between $3^{\circ}$ and $40^{\circ}$. The first disk is located 2 cm downstream from the interaction point, the last disk 23 cm. The first four layers consist of pixel detectors, whereas the last two layers are made up of both pixel and strip detectors.

The MVD is designed to provide a vertex resolution of 100 $\mu$m in the beam direction and a few tens of $\mu$m in the radial direction. Combined with the STT and GEM detectors, PANDA will achieve a momentum resolution of $\sigma_p / p < 1 \%$ \cite{mvd_tdr}.

\subsection{Straw Tube Tracker}


The STT is the main tracking detector of the target spectrometer and surrounds the MVD. Being inside a solenoid field, charged particles traverse it along helical trajectories. The STT is designed for particles in a momentum range from a few hundred MeV/c up to 8 GeV/c and a scattering angle range between $22^{\circ}$ and $140^{\circ}$. 

\begin{figure}
  \centering
  \includegraphics[width=\linewidth]{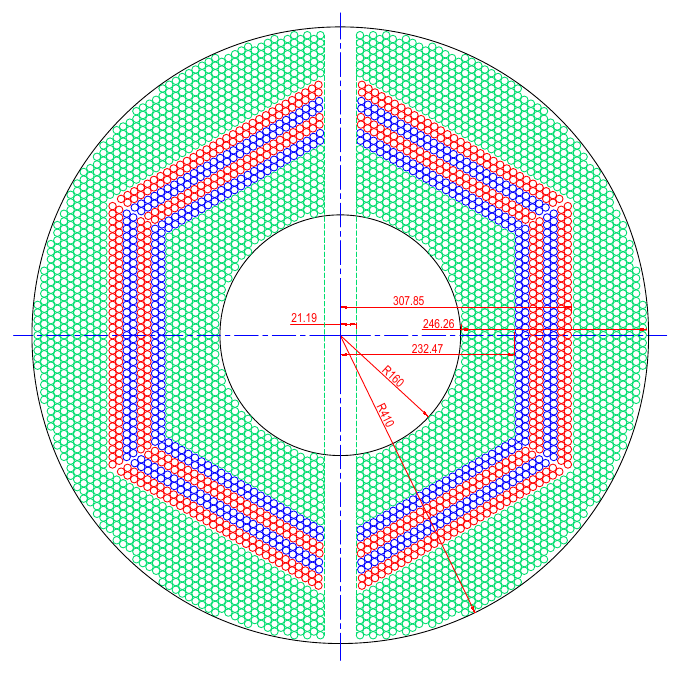}
  \caption{Cross sectional view of the Straw Tube Tracker in the $xy$-plane. The green circles represent parallel straws, the blue and red circles the skewed straws.}
  \label{fig:stt_layout}
\end{figure}

The detector consists of 4224 gas-filled tubes, arranged in a hexagonal layout and filling a cylindrical volume. A vertical gap is left at the centre to make room for the target system. The inner and outer tube layers are oriented parallel to the beam pipe, whereas the 8 intermediate layers are tilted by $\pm 2.9^{\circ}$. The purpose of the latter is to provide longitudinal information about the particle trajectories. A cross section of the STT is shown in Fig.~\ref{fig:stt_layout}, where the parallel tubes are presented in green and the skewed tubes in red or blue.

The tubes are made from a $27 \mu$m thick mylar foil with 10 mm in diameter and filled with a 90/10 gas mixture of Ar/CO$_2$. The tubes have a conductive inner layer and a 20 $\mu$m thick gold-plated tungsten anode wire at the centre. A potential differential of several kV is applied between the cathode layer and the anode wire. Charged particles traversing the tube ionize the gas, producing free ions and electrons that are accelerated towards the anode wire. In the vicinity of the wire ($\mathcal{O}(10 \, \mu \textrm{m})$) they produce more free electrons and ions in an avalanche-like manner, which in turn results in a signal, large enough for readout.

\begin{figure}
  \centering
  \includegraphics[width=\linewidth]{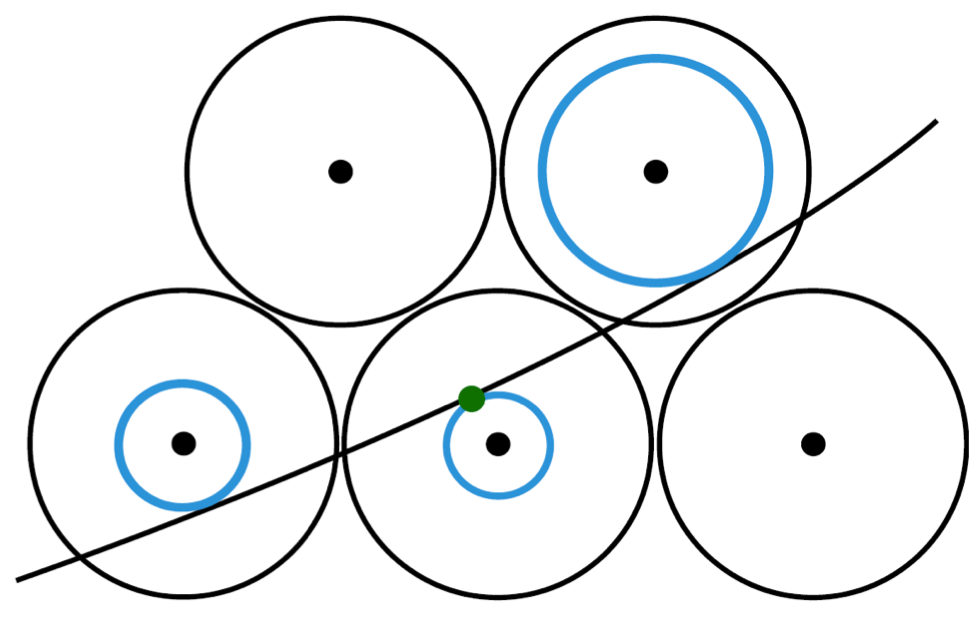}
  \caption{An illustration of a particle traversing three straw tubes, shown in the $xy$-plane. The isochrones are represented by blue circles. The green point indicates the correct position where the distance to the wire is minimal and the track is tangential to the isochrone.}
  \label{fig:isochrone}
\end{figure}

The time it takes the electrons to reach the anode wire is called the drift time. By measuring the shortest drift time within a signal, the minimal distance between the wire and the charged track can be inferred. This distance is usually referred to as the \textit{isochrone radius}, which describes a cylindrical surface around the wire and contains all possible positions that the particle might have traversed. This is illustrated in Fig.~\ref{fig:isochrone}. 

After a series of straw tubes has been associated to a common track candidate, the particle track is  reconstructed by fitting a trajectory such that it is tangential to the isochrones.
The resolution of the isochrone radius will be $< 150 \, \mu$m, whereas the resolution of $z$-position along the wire will be about 3 mm \cite{stt_tdr}.


\section{Pattern recognition}
\label{sec:pattern_recognition}

Track reconstruction begins with pattern recognition. In this phase, signals from individual detector elements are associated to a common track, originating from the traversing particle. 
The output consists of sets of hits together with an initial estimate for the track parameters, \textit{e.g.} based on a simple helix fit. The final track parameters are obtained later using a more sophisticated, usually adaptive fitting method that takes into account physical effects such as energy loss, as well as imperfections in the detector. However, at the pattern recognition stage, it is often not only convenient but also necessary to disregard these effects and instead use the helix shape as an approximation to the particle trajectory.

\subsection{Track representation}
\label{sec:track_representation}

Assuming that the trajectory of a particle with charge $Q$ moving in the solenoid field of the target spectrometer can be approximated by a helix or radius $r$, five parameters are needed to describe the track. A sixth parameter can be added to define a point along the helix trajectory.

The PANDA target spectrometer has a cylindrical design. A superconducting solenoid provides an approximately homogeneous magnetic field for the whole detector volume, bending the trajectories of charged particles into helices around an axis parallel to the beam axis. Choosing the $z$-axis along the beam direction, the track can be explicitly represented by the following five parameters:
\begin{itemize}
  \item $x_c$ - horizontal coordinate of the helix centre
  \item $y_c$ - vertical coordinate of the helix centre
  \item $\omega = Q/r$ - signed curvature of the helix
  \item $\phi_0$ - azimuthal angle at the reference $z_0$
  \item $\tan \lambda$ - dip angle of the helix with $\lambda = \arctan p_z/p_t$
\end{itemize}
In this representation, the helix can be analytically described in two projections. In the $xy$-projection the helix becomes a circle, defined as
\begin{equation}
  (x - x_c)^2 + (y - y_c)^2 = r^2.
\end{equation}
It is reduced to a line in the $z\phi$-projection, \textit{i.e.}
\begin{equation}
  \phi(z) = kz + \phi_0,
\end{equation}
with the slope parameter $k \equiv \tan \lambda$. The azimuthal angle $\phi$ can be replaced by the arc length
\begin{equation}
  S = (\phi_0 - \phi) rQ,
\end{equation}
where $Q$ is the charge of the track. The arc length is defined such that it is zero at the starting point of the track and increases positively, regardless of the charge. The linear relation then becomes
\begin{equation}
  z(S) = \tan \lambda S + z_0.
  \label{eq:z_of_S}
\end{equation}
Eq.~(\ref{eq:z_of_S}) simplifies the problem considerably, and the task reduces to identifying a line among a set of points in the $Sz$-space.




\section{Longitudinal track reconstruction}
\label{sec:longitudinal_track_reconstruction}

\begin{figure}
  \centering
  \includegraphics[width=\linewidth]{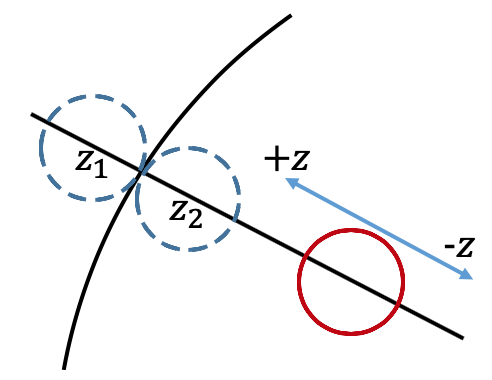}
  \caption{The isochrone alignment procedure. The isochrone ellipse (red) in the $xy$-projection is aligned along the STT tube projection such that it is tangential to the track (black curve). Two possible solutions for the $z$-position are obtained, $z_1$ and $z_2$.}
  \label{fig:alignment}
\end{figure}

The PANDA software framework, \textit{PandaRoot} \cite{pandaroot}, contains a variety of global and local track finding algorithms. While the global methods aim to take the whole detector into account, the local methods focus on one or few sub-detectors, such as the Straw Tube Tracker. For the STT in particular, the existing local track finders focus on reconstructing the transversal track components. To supplement these track finders, a set of algorithms dedicated to the extraction of longitudinal position and momentum information of track candidates was developed and named the \textit{PzFinder} after the third momentum component.
Since the \textit{PzFinder} is used during the early stage of the track reconstruction, the particle trajectory is approximated by a helix which is described as in Sec.~\ref{sec:track_representation}. The \textit{PzFinder} requires the three parameters $x_c, y_c$, and $Q/r$ to be known and provides estimates for $\phi_0$ and $\tan \lambda$ in return, yielding a complete description of the track.

Using this approximation, the track becomes a circle in the $xy$-projection with constant transversal and longitudinal momentum components $p_t$ and $p_z$, respectively. Hence, the particle traverses at a constant rate in $z$ with respect to the helix axis. According to Eq.~(\ref{eq:z_of_S}), all true hit positions of a track candidate should then lie on a straight line in the $(S,z)$ coordinate space.

The $z$-position for each hit in a skewed STT tube is extracted through an isochrone alignment procedure, illustrated in Fig.~\ref{fig:alignment}. 
The isochrone radius for each hit is determined beforehand outside the \textit{PzFinder}. 
Since these tubes are skewed, the projection of the isochrone onto the $xy$-plane becomes an ellipse. 
The isochrone is aligned such that its centre position lies along the wire and is tangential to the particle trajectory.
This alignment implicitly gives a solution for the $z$-position.
However, during the alignment procedure, two solutions/points are obtained, introducing a left/right ambiguity with one solution on either side of the trajectory.

In the following, three different methods are presented to resolve this ambiguity.

\subsection{Combinatorial path finder}
\label{sec:combinatorial_path_finder}

\begin{figure}
  \centering
  \subfloat[]{\label{subfig:combi1} \includegraphics[width=0.49\linewidth]{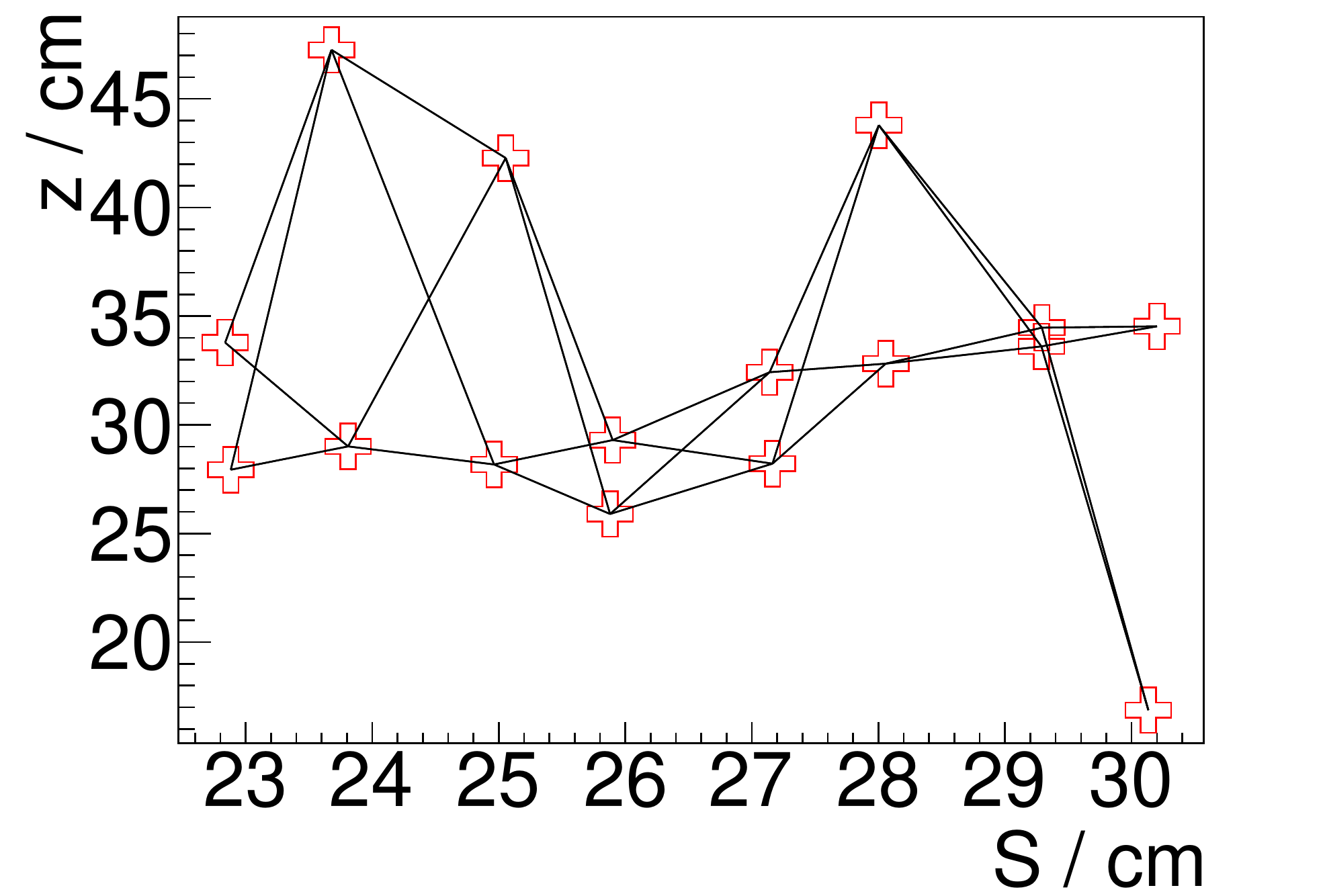}}
  \subfloat[]{\label{subfig:combi3}  \includegraphics[width=0.49\linewidth]{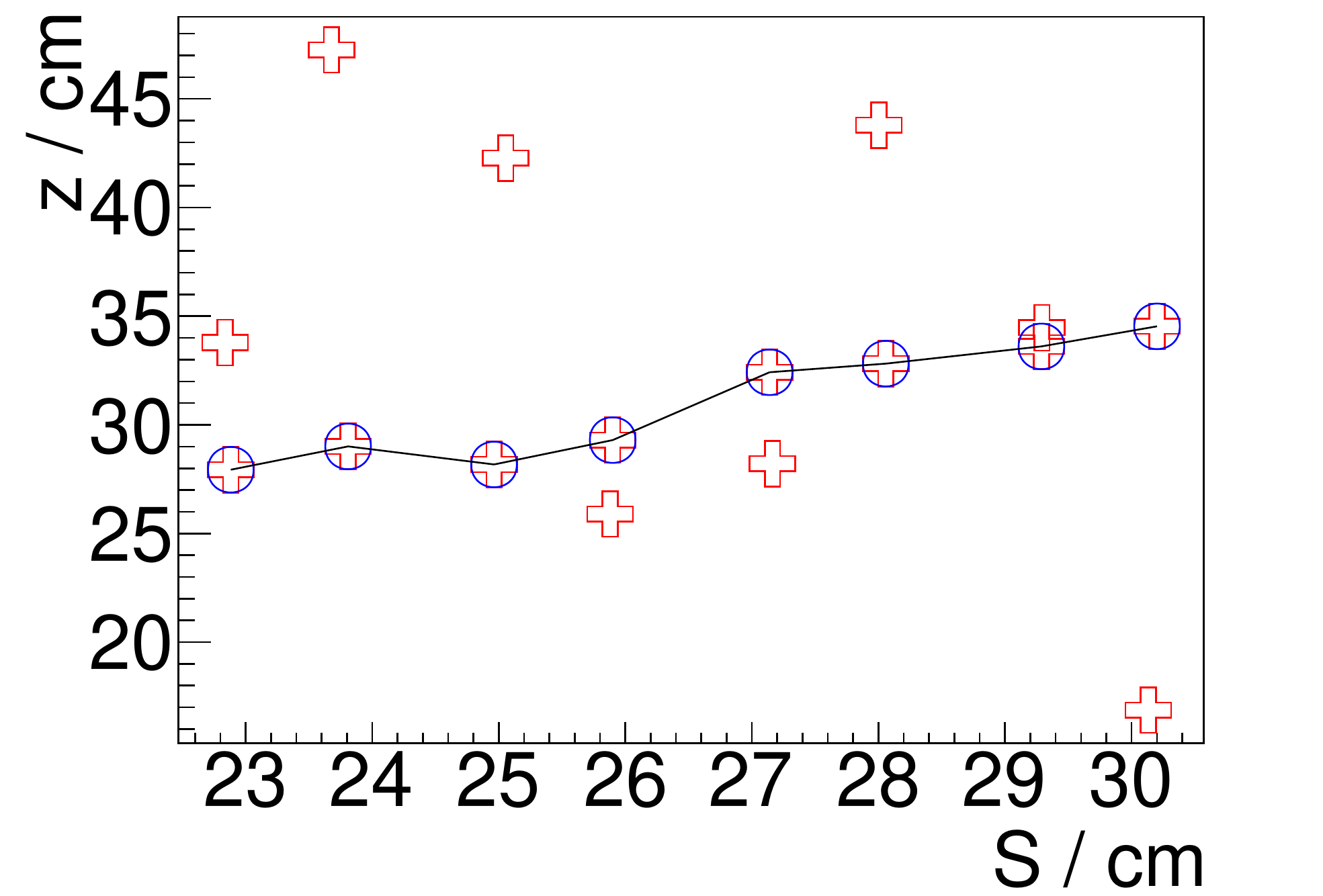}}
  \caption{A visualisation of the combinatorial path finder. \ref{sub@subfig:combi1} All possible connections are considered. The black lines are generated by the algorithm to calculate angles. The encircled points are the  $(S,z)$ points that were selected as part of the straightest path. \ref{sub@subfig:combi3} Only the lines corresponding to the found path remain.}
  \label{fig:path_finder}
\end{figure}

A combinatorial approach was developed to find the "straightest" path from one end of the track to the other, similar to a track road procedure \cite{tracking_overview}.
First, all possible connections between the $(S,z)$ solutions of neighbouring STT tubes are determined.
With two $(S,z)$ solutions for each hit, this yields four possible connections between neighbouring tubes.
Fig.~\ref{fig:path_finder} illustrates the procedure.
Given a track candidate with $N$ hits in the skewed STT layers, the total number of generated lines is $4(N-1)$. 
Since there are eight layers of skewed STT tubes, a typical track passing fairly straight through all these layers will have 24 connections to be considered.

The "straightest" path is determined by evaluating the cost function
\begin{equation}
  w = \sum^{n}_{j} (180^{\circ} - \theta_j)^2,
  \label{eq:cost_function}
\end{equation}
where $\theta_j$ is the opening angle between two lines connecting at the same $(S,z)$ point. With two possible points for each STT hit, this leaves $2^N$ possible paths to be evaluated. To reduce the computational footprint, paths containing an opening angle smaller than $90^{\circ}$ are discarded right away. Among the remaining points, the correct ones are chosen by minimising the cost function Eq.~(\ref{eq:cost_function}). In Fig.~\ref{fig:path_finder}, the selected positions are marked by blue circles. For this example track, these positions are also the ones that are closest to the true positions, marked by green stars. This indicates that the selection process was successful.


The combinatorial nature of this approach brings about a possibly high computational cost. Since the STT has eight layers of skewed tubes, any fairly straight particle trajectory creates eight or nine skewed STT hits, resulting in 256 or 512 combinations to be evaluated. Tracks with a higher curvature can leave even more hits, increasing the number of combinations exponentially. Furthermore, the algorithm assumes that all skewed STT hits have been correctly associated with the track by the preceding pattern recognition.

These considerations leave room for other options to be explored.

\subsection{Hough transformation}
\label{sec:hough_transformation}

The Hough transformation \cite{hough} has been a staple in track reconstruction in particle physics for more than half a century. It was designed to detect simple shapes, originally lines, but later also more complex one such as circles or ellipses \cite{hough_more}. Since the parameterisation of these shapes becomes part of the algorithm, it performs best when the points follow this parameterisation closely and the shape is not distorted by secondary effects such as energy loss. Since the problem at hand is about finding points on the same line in $(S,z)$ space, it is reasonable to expect that the classical Hough transformation could be a good solution.

\begin{figure}[ht]
  \begin{minipage}{.49\linewidth}
  \centering
  \includegraphics[width=\linewidth]{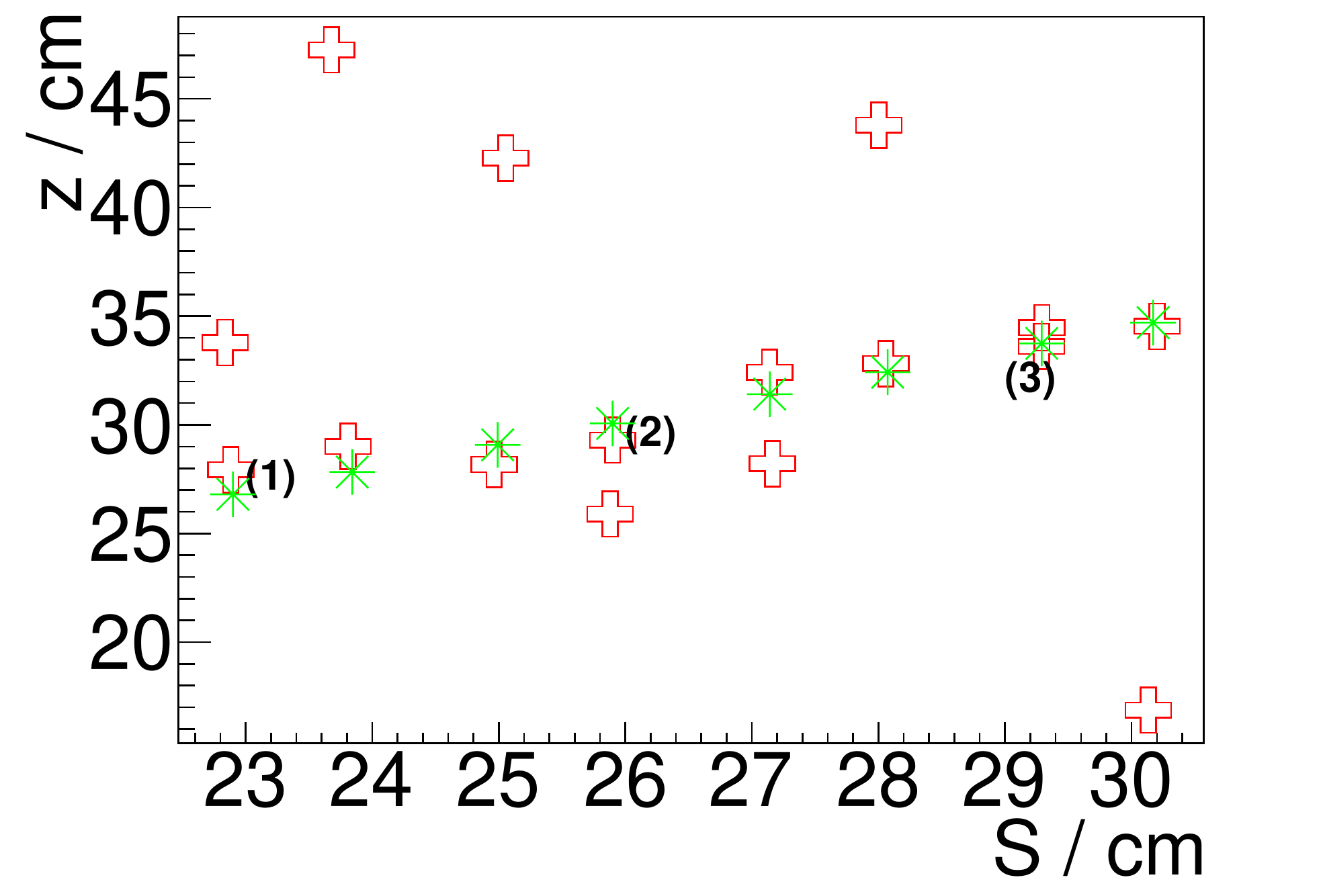}\\
  \textbf{(a)}\\
  \includegraphics[width=\linewidth]{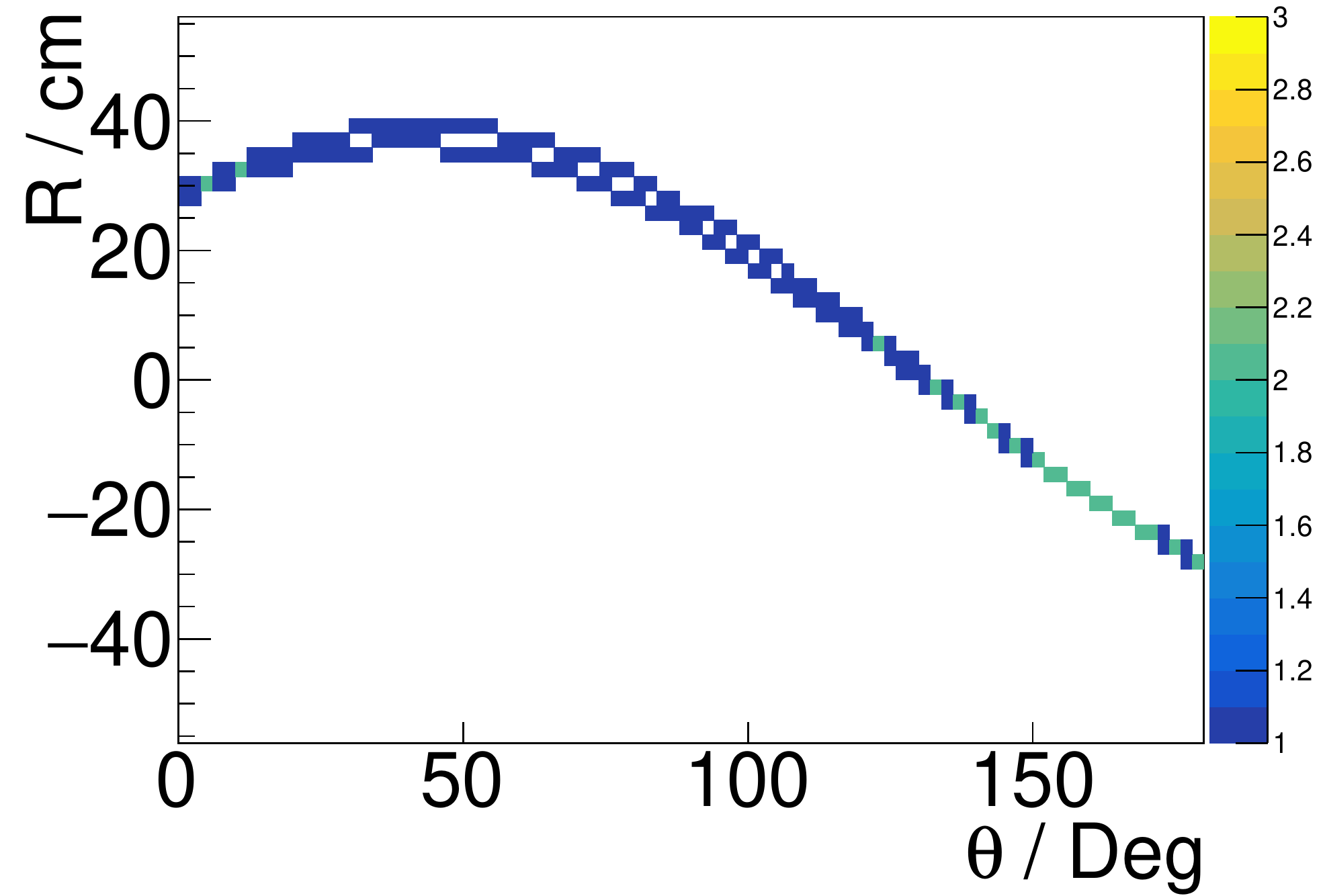}\\
  \textbf{(c)}
  \end{minipage} 
  \begin{minipage}{.49\linewidth}
  \centering
  \includegraphics[width=\linewidth]{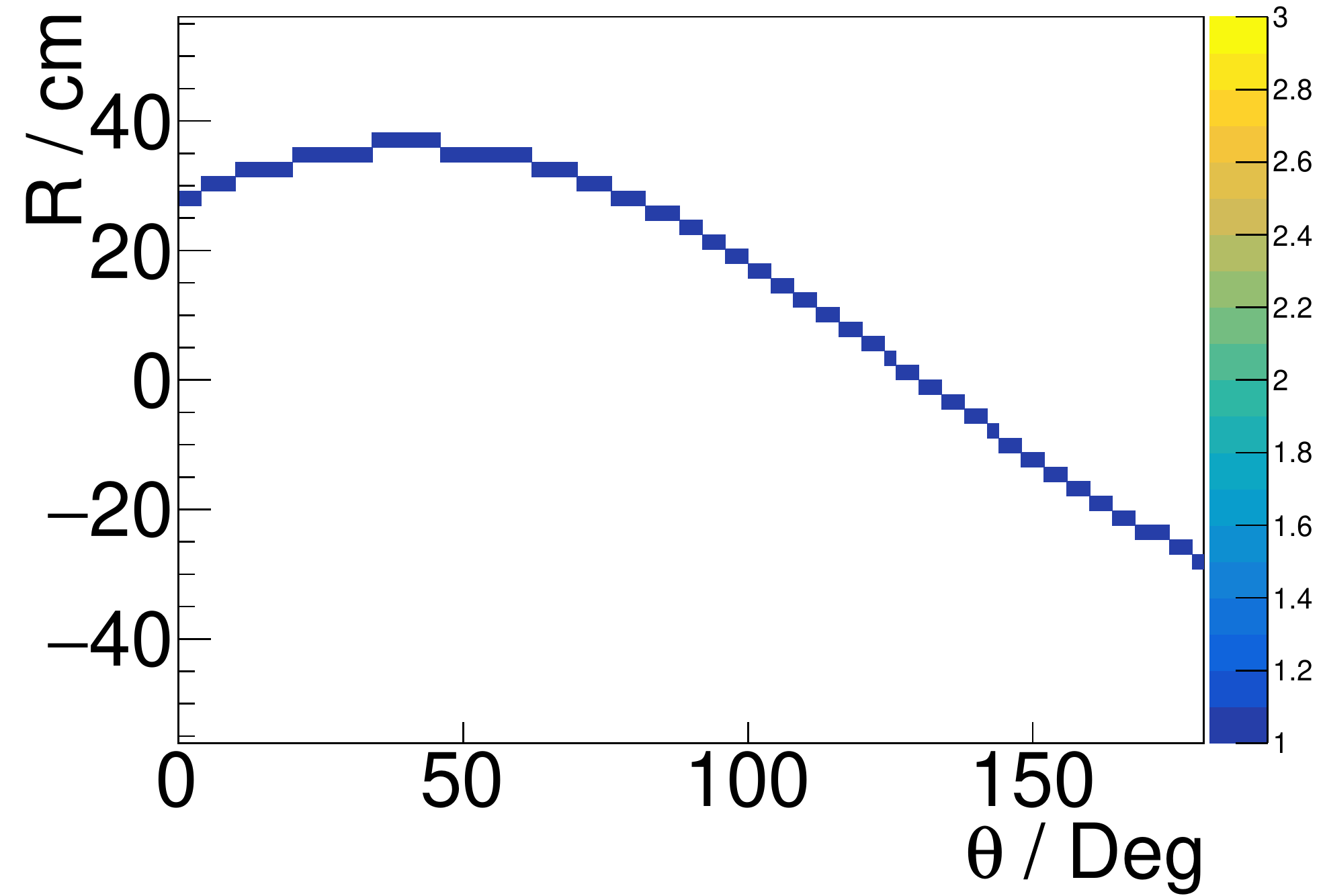}\\
  \textbf{(b)}\\
  \includegraphics[width=\linewidth]{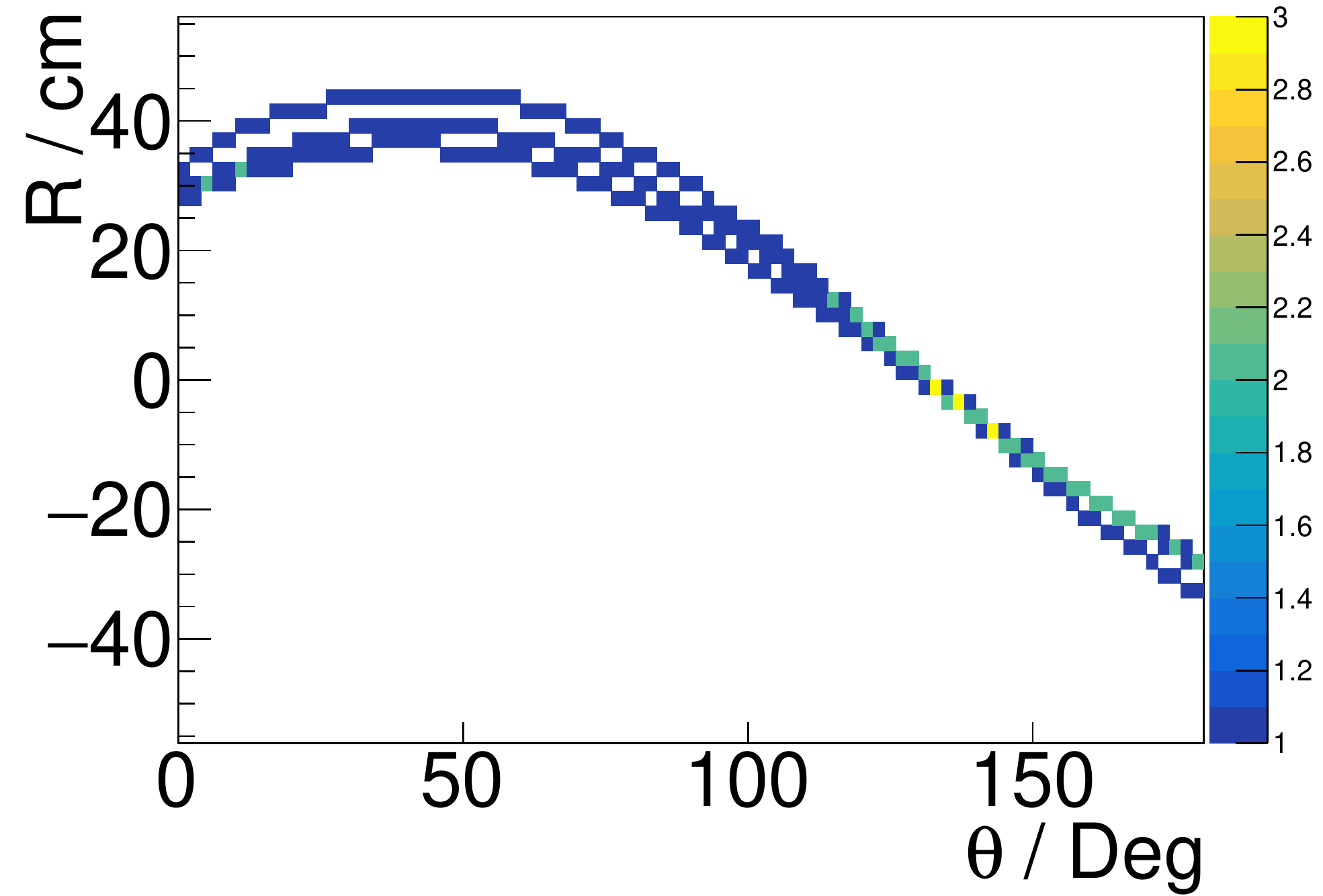}\\
  \textbf{(d)}
  \end{minipage} 
  \caption{A visualisation of the Hough transformation algorithm. \textbf{(a)} $(S,z)$ points with three example hits highlighted. \textbf{(b)}, \textbf{(c)} and \textbf{(d)}, the accumulator space after the line parameters of hits 1, 2 and 3 have been added, respectively.}
  \label{fig:hough}
\end{figure}

In order to avoid unbound parameters, the line is described using the Hesse normal form with the closest distance to the origin $R$ and the inclination angle $\theta$:
\begin{equation}
  R = z \cos\theta + S \sin\theta
  \label{eq:hesse_normal_form}
\end{equation}
In the Hough transformation a set of lines is generated for each point. All lines go through the point, but at different inclination angles. The line parameters are then filled into an accumulator space that is spanned by $\theta$ and $R$. In the accumulator space, local maxima will emerge for line parameters that connect multiple points. In our case, the algorithm searches for only one line connecting the largest number of points. Hence, the corresponding parameters can be found by selecting the global maximum in the accumulator space. The $(S,z)$ points contributing to this maximum are chosen as the correct ones. The procedure is illustrated in Fig.~\ref{fig:hough}. In case the global maximum is ambiguous, \textit{i.e.} two bins with the maximum number of entries are present, a line fit for both sets of $(S,z)$ hits is performed and the bin resulting in the smallest $\chi^2$ is selected as the correct one.

The algorithm can incorporate hits from the MVD detector to improve the procedure. However, these have to be supplied by a separate, local pattern recognition algorithm. The method can be extended to reject false MVD hits as well.

\subsection{Recursive annealing fit}
\label{sec:recursive_annealing_fit}

Annealing procedures found their way into track reconstruction several decades ago \cite{tracking_overview}. One possible application is to remove wrongly matched hits from a track and thereby improve the track fit \cite{annealing_belleii}.
Thus, the recursive annealing fit operates by removing outliers within a set of measurements. The rejection of these outlying measurements can be based on the measurement of a residual, here the distance between one $(S,z)$ point and a line fit to all $(S,z)$ points.

The initial line fit is performed to all $(S,z)$ points in a track candidate. Since the detector resolution is much better in the transversal direction than in the longitudinal, only the uncertainty of the $z$-component of the position is taken into account. Given that an iterative minimisation has to be performed, this is a helpful simplification of the fit. The line expression in Eq.~(\ref{eq:z_of_S}) is fitted to the points to extract the slope $\tan \lambda$ and the starting position $z_0$.
The fit minimises the $\chi^2$
\begin{equation}
  \chi^2 = \sum_i^n \frac{(z_i - kS_i - z_0)}{\sigma_i^2}
\end{equation}
with respect to the slope $k$ and the intercept $z_0$. The index $i$ runs over all $(S,z)$ points and the uncertainty $\sigma_z$ of the  $i$-th hit is denoted $\sigma_i$.

The $(S,z)$ points with the largest contribution to the $\chi^2$ are rejected from the skewed STT hits after the initial fit. Specifically, the $(S,z)$ point with the largest residual is removed and the remaining points are fitted anew. This procedure is repeated until one $(S,z)$ point has been removed for every skewed STT hit.


\section{Data samples}
\label{sec:data_samples}

To evaluate the algorithms, two different sets of data samples were produced.

First, a particle generator was used to simulate 1000 events with four muons, two of each charge, orginating from the interaction point and isotropically and monochromatically distributed. This was done with each of the momenta of 0.1, 0.2, 0.5, 1.0, 2.0, and 5.0 GeV/c to cover a wide range of expected particle momenta. Muons have two main advantages: i) they interact mainly electromagnetically with the detector material, hence their measurement is less obscured by strong processes compared to hadrons and ii) being 200 times heavier than electrons, they are less prone to change direction through scattering. The muons in these samples originate from the interaction point and are emitted uniformly within the acceptance of the Straw Tube Tracker, \textit{i.e.} the polar angle $\theta \in [22^\circ, 140^\circ]$ and the azimuthal angle $\phi \in [0^\circ, 360^\circ]$.

More general data sets are provided by the Dual Parton Model generator \cite{dpm}, offering a realistic composition of known processes that occur in $\overline{p} p$ collisions. Four samples at the beam momenta 1.642, 4.6, 7.0, and 15.0 GeV/c, with 5000 events in each sample, were produced in this study. The first three points bear relevance to specific physics channels, whereas the fourth point corresponds to the maximum beam momentum at which the PANDA experiment will operate.

\section{Results}
\label{sec:results}

\subsection{Efficienty and Purity}
\label{sec:efficiency_and_purity}

\begin{figure*}
  \centering
  \subfloat[]{\label{subfig:pzperf_eff} \includegraphics[width=0.49\linewidth]{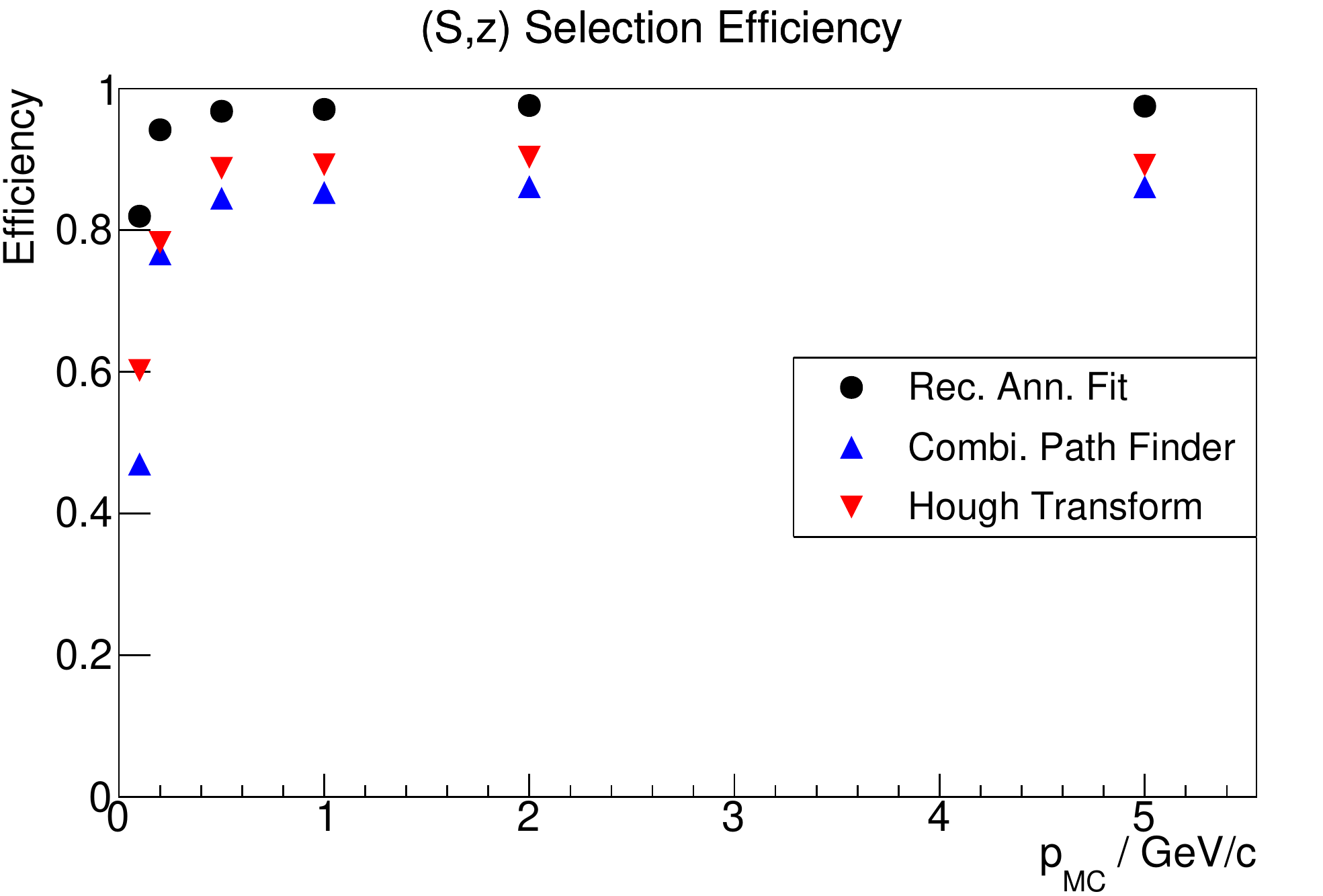}}
  \subfloat[]{\label{subfig:pzperf_pur} \includegraphics[width=0.49\linewidth]{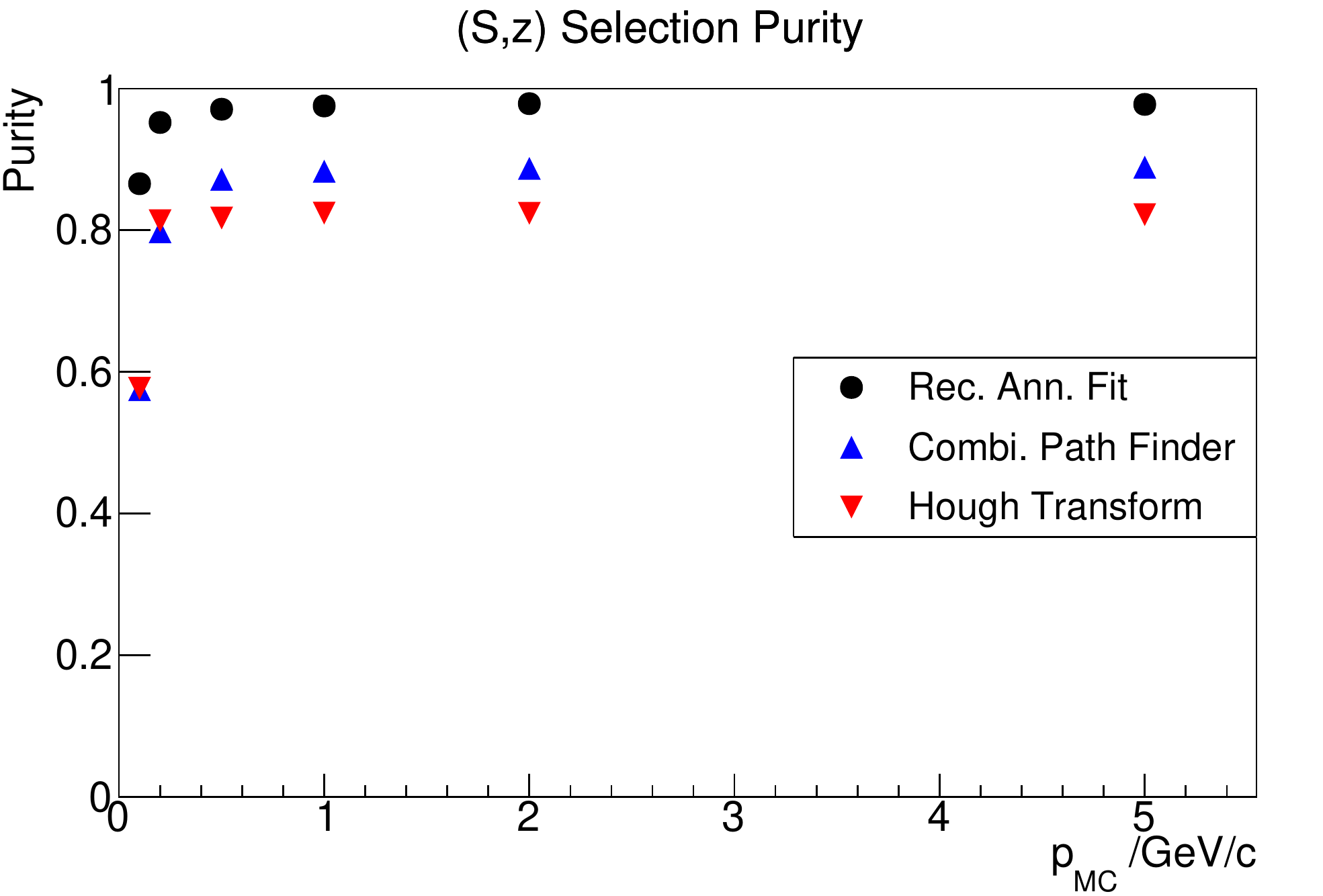}}
  \caption{Sample efficiency \ref{sub@subfig:pzperf_eff} and purity \ref{sub@subfig:pzperf_pur} of $(S,z)$ points selected by the Recursive Annealing fit (black dots), Combinatorial Path Finder (blue upwards triangles), and Hough transformation (red downwards triangles). The statistical uncertainties have been calculated, but vanish behind the markers in the plot.}
  \label{fig:pzperf}
\end{figure*}

As a first step, the performance of the \textit{PzFinder} is assessed by investigating the efficiency and purity of the $(S,z)$ point selection, \textit{i.e.} the resolution resulting from the left-right ambiguity. It is assumed that for each skewed STT hit in a track candidate one correct and one false $(S,z)$ point exists. The $(S,z)$ point closest to in $z$ to the Monte Carlo truth position is then selected as the correct one.
The selection efficiency is defined as
\begin{equation}
    \epsilon = \frac{N_{hit}}{N_{hit}^{MC}},
\end{equation}
where $N_{hit}$ is the number of found hits that belong to one reference track and $N_{hit}^{MC}$ the total number of hits created by that reference track. Introducing the total number of found hits as $N_{hit}^{tot}$, the purity is given as
\begin{equation}
    purity = \frac{N_{hit}}{N_{hit}^{tot}}.
\end{equation}

The efficiencies and purities of the three algorithms are shown in Fig.~\ref{fig:pzperf} for the momentum points described in Sec.~\ref{sec:data_samples}. While all three algorithms reach $80 \%$ for momenta above 0.2 GeV/c in both observables, the Recursive Annealing Fit performs best with efficiencies and purities above $95 \%$.

\subsection{Hough transformation binning}

Naturally, the bin width of $R$ and $\theta$ can have great impact on the efficiency of this procedure. The choice of a proper bin width is often informed by external parameters, such as the detector resolution.

\begin{figure}
  \centering
  \includegraphics[width=\linewidth]{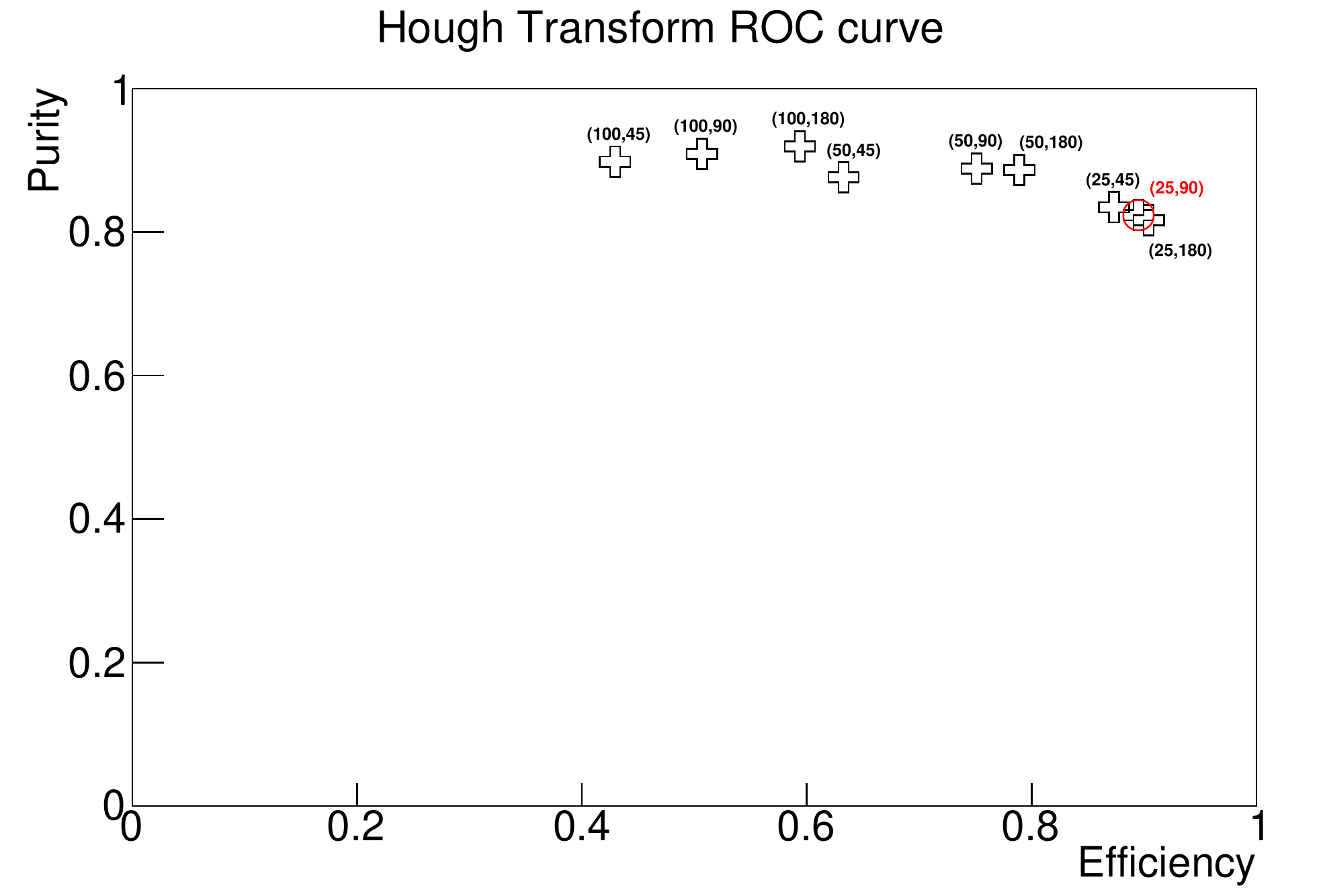}
  \caption{RoC curve of the Hough transformation algorithm with different bin widths of the accumulator space. Each point is classified by the number of bins in $R$ and $\theta$. The optimal choice is highlighted in red.}
  \label{fig:houghroc}
\end{figure}

To find the optimal binning for sec.~\ref{sec:efficiency_and_purity}, different numbers of bins were tested on a sample of 1 GeV/c muons: $N^R_{bin} = 25,50,100$ and $N^{\theta} = 45,90,180$ for the parameters $R$ and $\theta$, respectively. Fig.~\ref{fig:houghroc} displays the purity and efficiency for these binnings as a pseudo \textit{Receiver Operating Characteristics curve} (ROC curve), illustrating the performance of binary classifiers.

The figure shows a good balance of binning choices, as some favour efficiency over purity and vice versa. The parameter set closest to (1,1) on the curve is chosen as optimal. Henceforth, the Hough transformation was used with the parameters $N^R_{bin} = 25$ and $N^{\theta}_{bin} = 90$. The optimal binning for this procedure depends on the detector resolution and thereby on the particle momentum. However, with further increasing particle momenta, the momentum resolution increases and hence this binning will remain a good choice.

\subsection{Spatial Resolution}
\label{sec:spatial_resolution}

\begin{figure*}
    \centering
    \subfloat[]{\label{subfig:z}
    \includegraphics[width=0.49\linewidth]{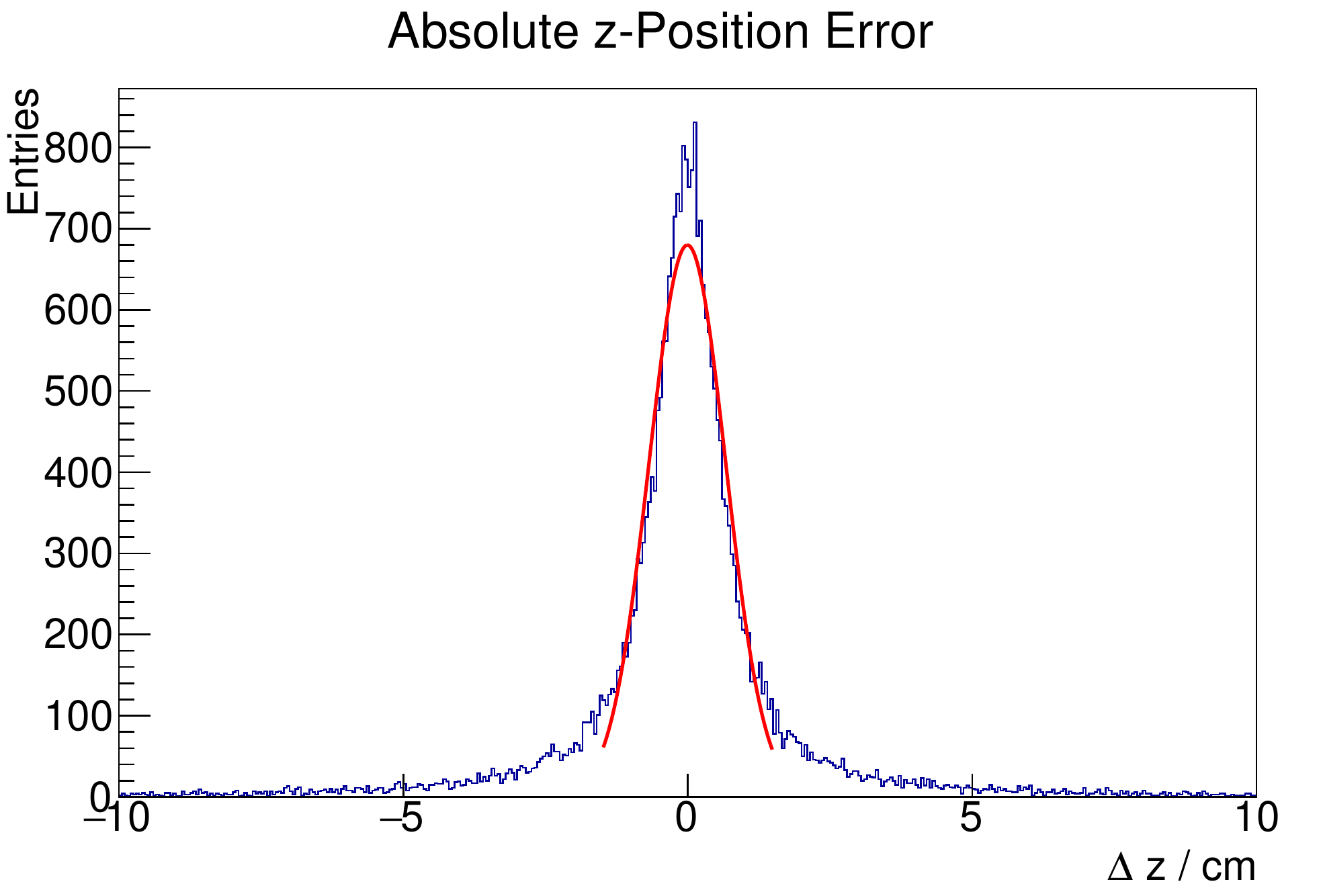}}
    \subfloat[]{\label{subfig:pl}
    \includegraphics[width=0.49\linewidth]{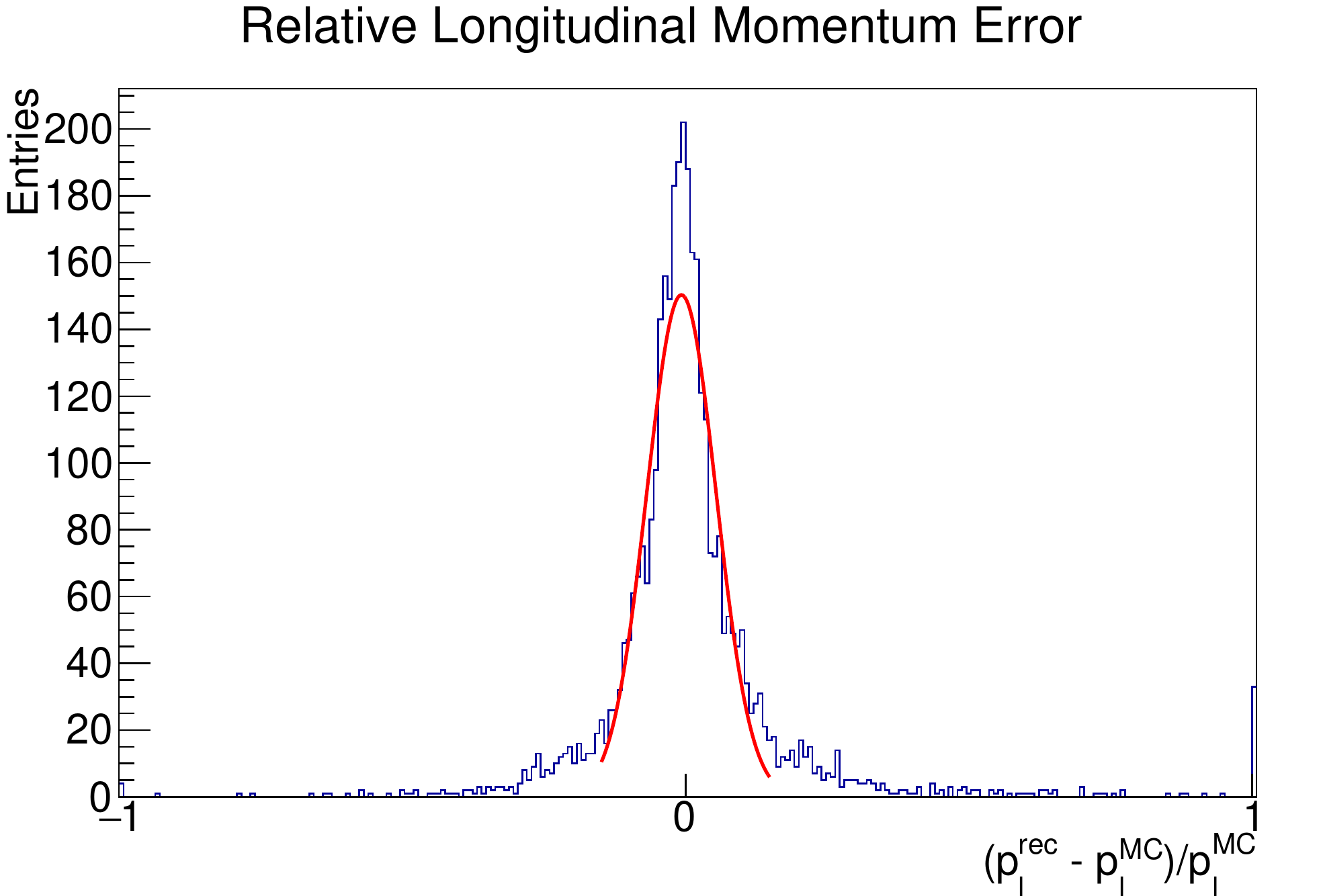}}
    \caption{Example distributions determining the distribution mean, root mean square, and standard deviation. A Gaussian function was fitted in both cases (red line). \ref{sub@subfig:z} The reconstructed longitudinal position $z$. \ref{sub@subfig:pl} The reconstructed longitudinal momentum $p_l$. The plots were produced from a data sample with muons with a momentum of 1 GeV/c.}
    \label{fig:resolution_determination}
\end{figure*}

\begin{figure*}
  \centering
  \subfloat[]{\label{subfig:deltazmean} \includegraphics[width=0.33\linewidth]{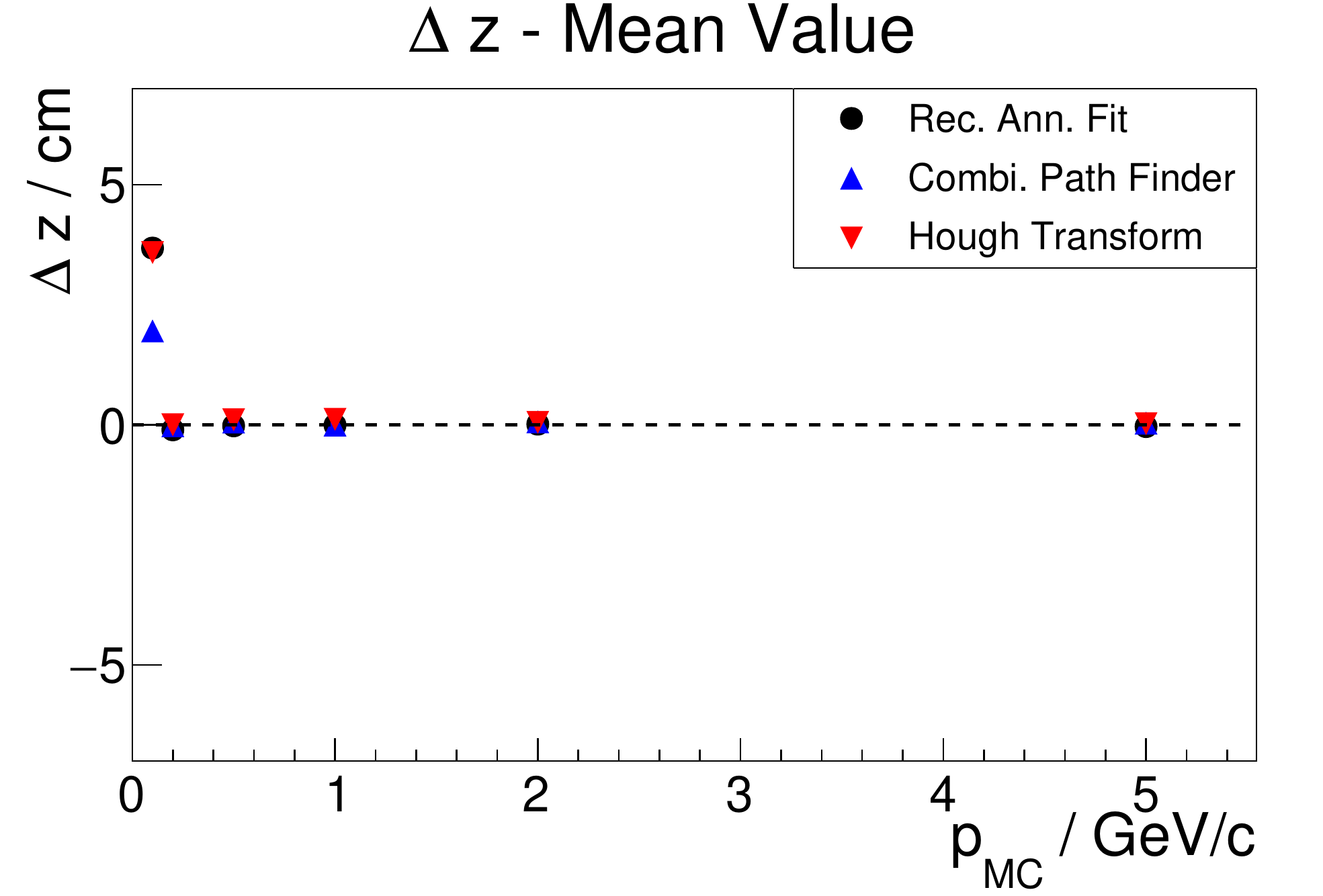}}
  \subfloat[]{\label{subfig:deltazrms}  \includegraphics[width=0.33\linewidth]{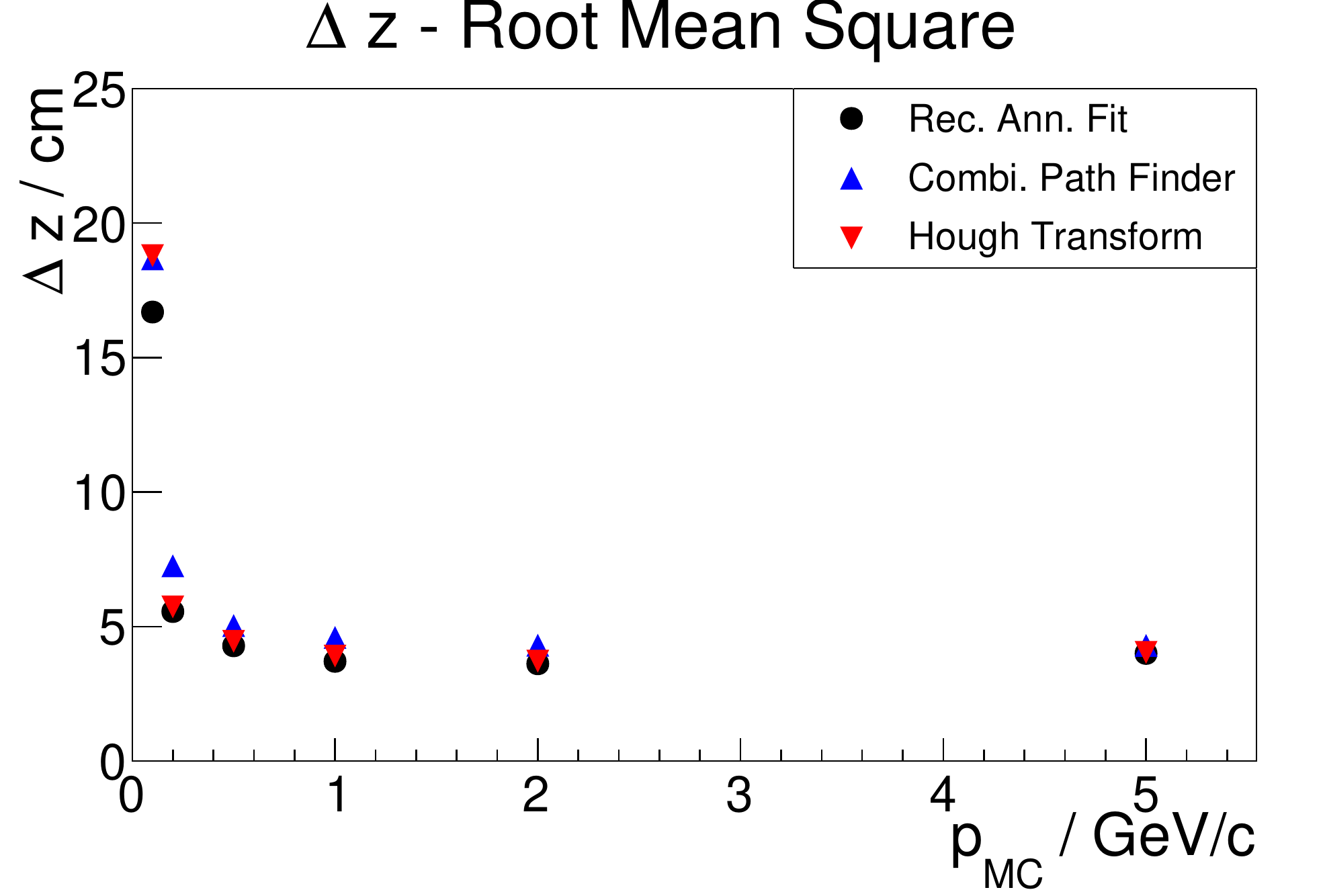}}
  \subfloat[]{\label{subfig:deltazstd}  \includegraphics[width=0.33\linewidth]{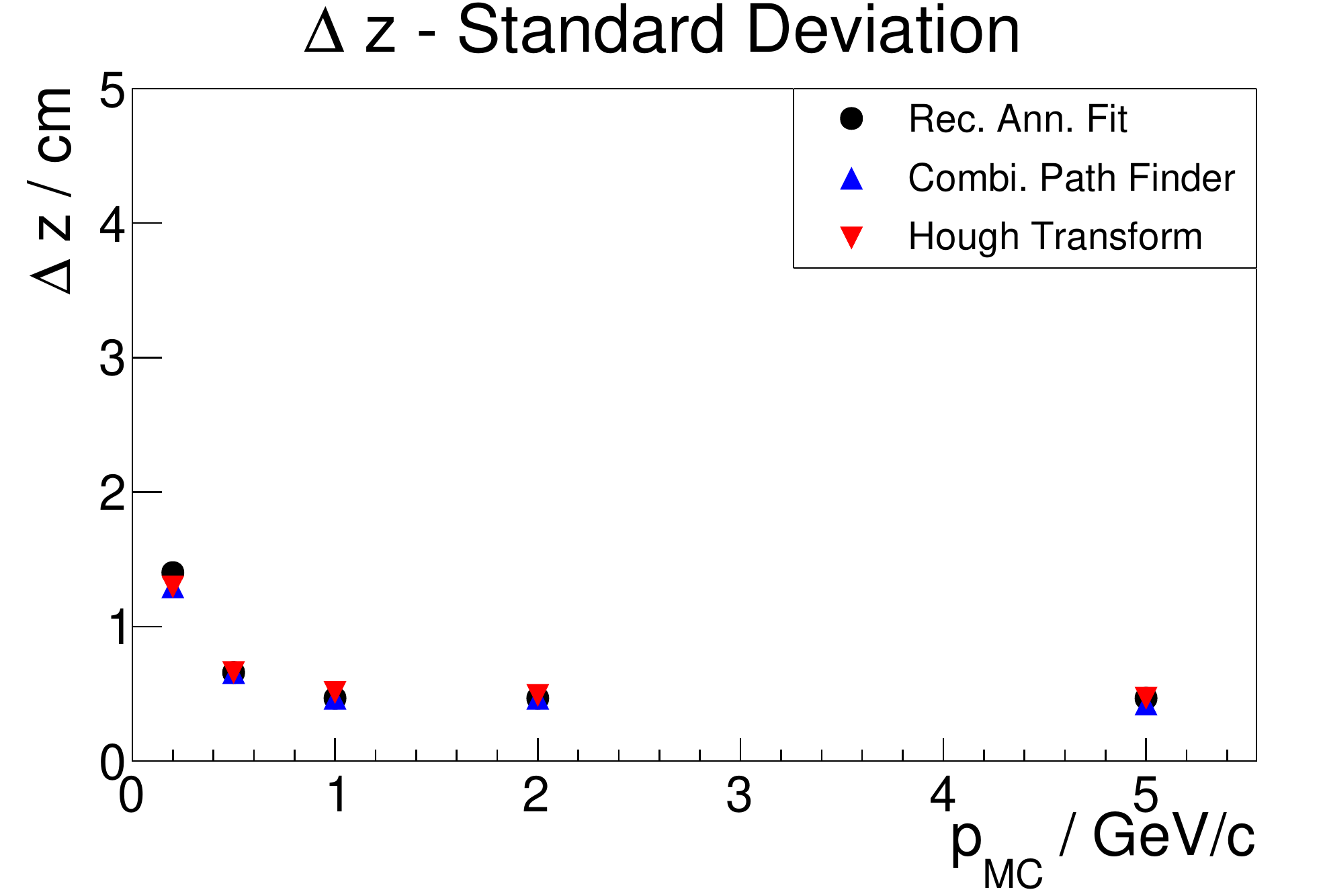}}
  \caption{Quality observables of the $z$-position of $(S,z)$ points selected by the Recursive Annealing fit (black dots), Combinatorial Path Finder (blue upwards triangles), and Hough transformation (red downwards triangles). Shown are distribution mean \ref{sub@subfig:deltazmean}, the root mean square \ref{sub@subfig:deltazrms}, and the standard deviation \ref{sub@subfig:deltazstd}.}
  \label{fig:deltaz}
\end{figure*}

\begin{figure*}
  \centering
  \subfloat[]{\label{subfig:plresmean} \includegraphics[width=0.33\linewidth]{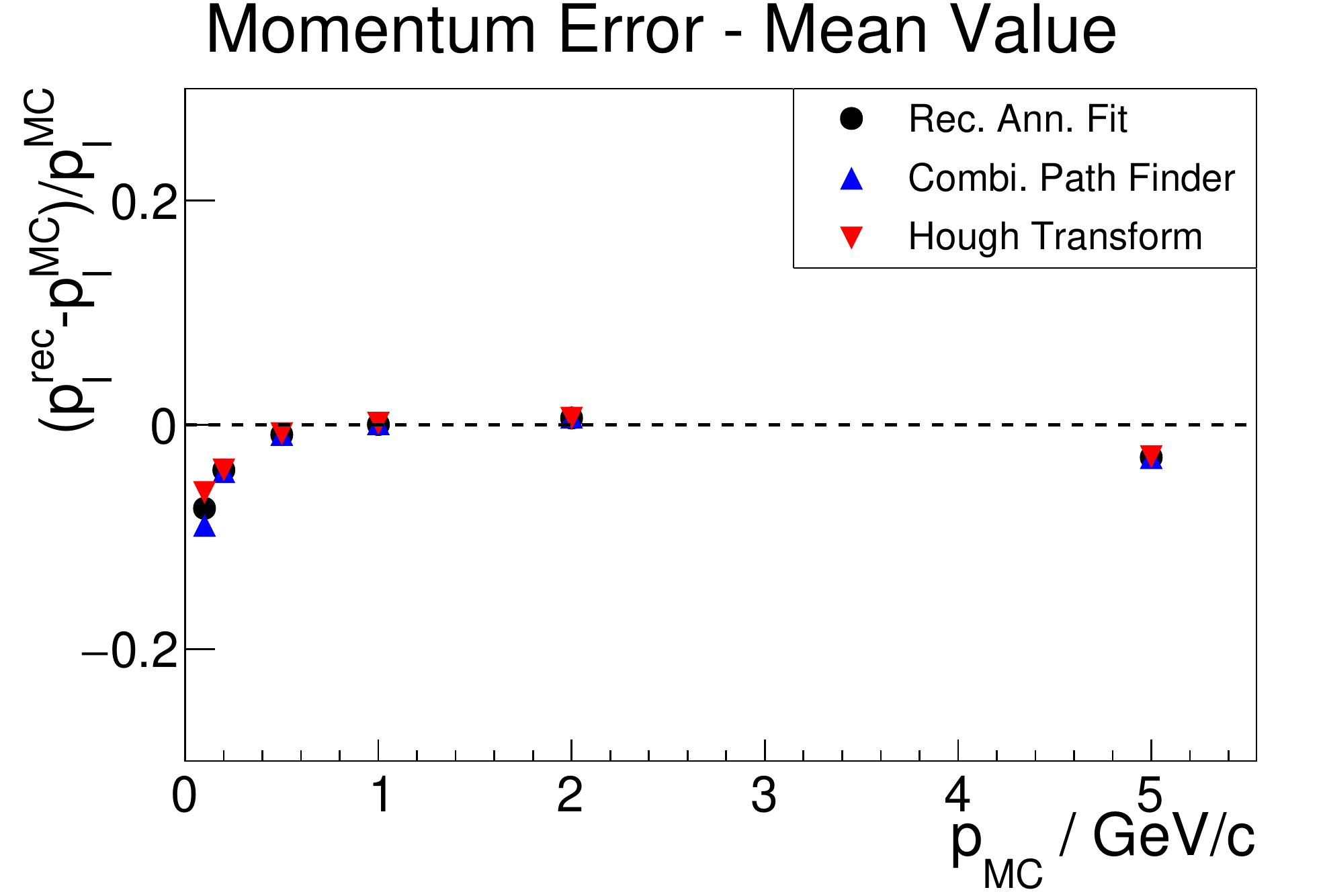}}
  \subfloat[]{\label{subfig:plresrms}  \includegraphics[width=0.33\linewidth]{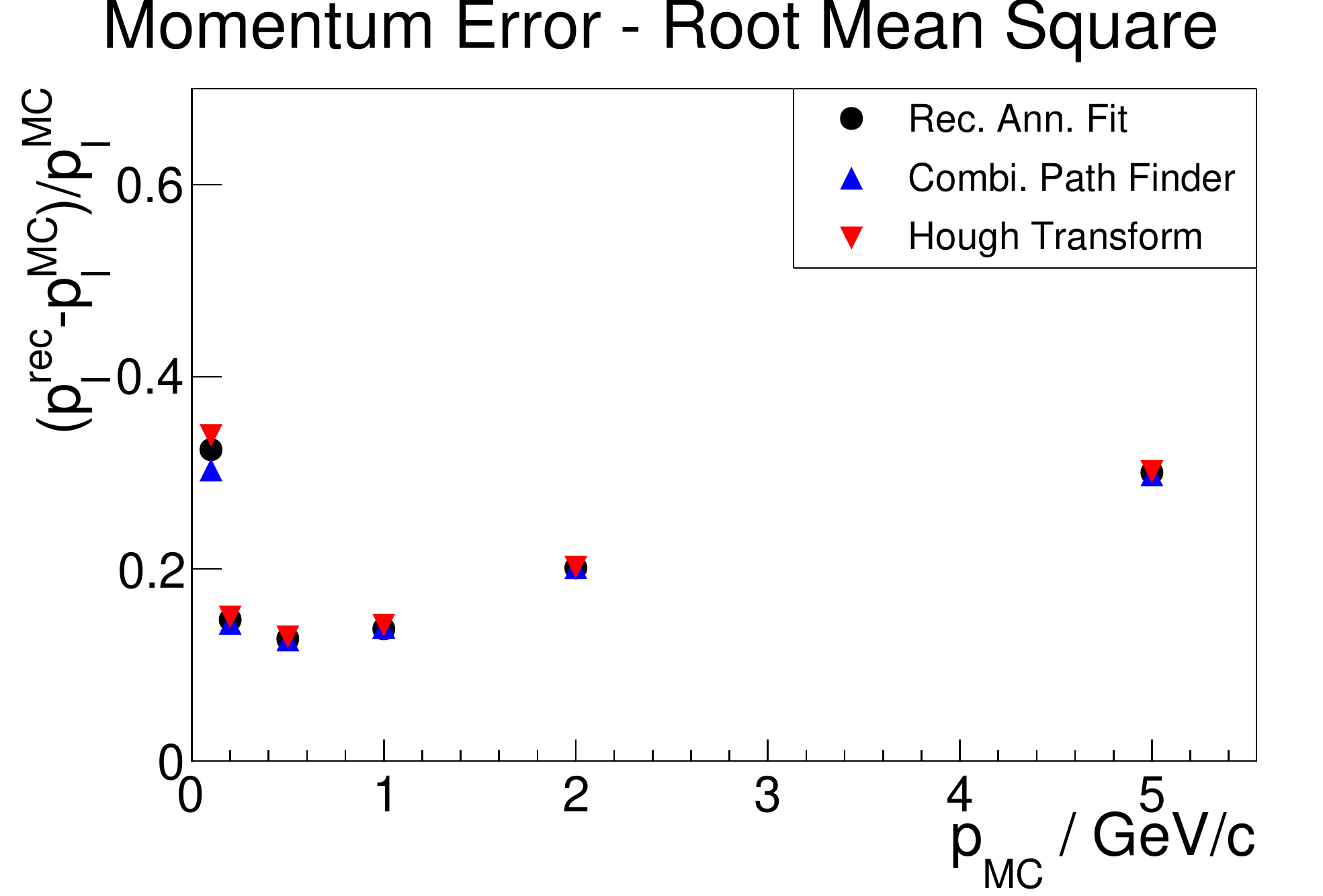}}
  \subfloat[]{\label{subfig:plresstd}  \includegraphics[width=0.33\linewidth]{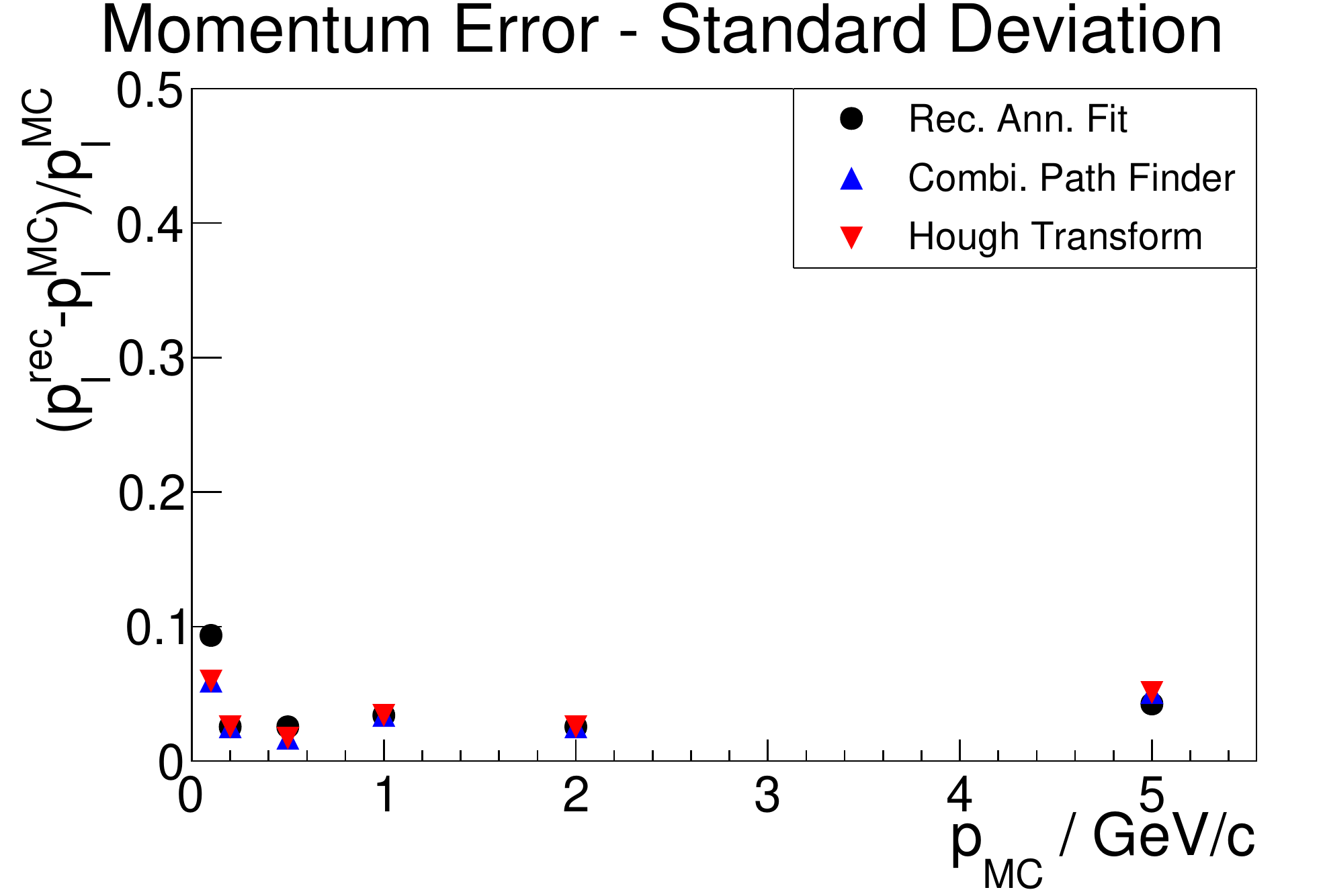}}
  \caption{Quality observables of the longitudinal momentum $p_l$ of track reconstructed by the Recursive Annealing fit (black dots), Combinatorial Path Finder (blue upwards triangles), and Hough transformation (red downwards triangles). Shown are distribution mean \ref{sub@subfig:plresmean}, the root mean square \ref{sub@subfig:plresrms}, and the standard deviation \ref{sub@subfig:plresstd}.}
  \label{fig:plres}
\end{figure*}

After selecting the $(S,z)$ points of a track candidate with the \textit{PzFinder}, the spatial reconstruction quality can be assessed by studying the $z$-position error \\$\Delta z = z_{rec} - z_{MC}$ for each hit. An example distribution for the reconstructed longitudinal position $z$ of 1 GeV/c muons is shown in Fig.~\ref{subfig:z}. Since the distribution does not always follow a well defined Gaussian shape, the standard deviation was approximated from the Full Width at Half Maximum (FWHM) using
\begin{equation}
\sigma_z = \textrm{FWHM}/2.355.    
\end{equation}
Fig.~\ref{fig:deltaz} shows the $\Delta z$ distribution mean, root mean square, and standard deviation for all three algorithms. Above particle momenta of $0.1 \,$GeV/c, the mean values of the distributions are close to zero. The root mean square values are around 3 cm for higher particle momenta. The standard deviation, on the other hand, quickly reaches values below 5 mm. Even before using an adaptive fitting method, \textit{e.g.} Kalman Filter \cite{kalman}, this comes close to the design value for the resolution of 3 mm \cite{stt_tdr}.

\subsection{Longitudinal Momentum Resolution}

To assess the quality of the longitudinal momentum reconstruction, we define the relative longitudinal momentum resolution as
\begin{equation}
    Res(p_l) = \frac{\left(p_l^{rec} - p_l^{MC}\right)}{p_l^{MC}},
\end{equation}
with $p_l^{rec}$ and $p_l^{MC}$ being the reconstructed and Monte Carlo momentum, respectively. An example distribution of the reconstructed longitudinal momentum $p_l$ for 1 GeV/c muons can be seen in fig.~\ref{subfig:pl}. The same considerations as in sec.~\ref{sec:spatial_resolution} also apply here.
The quality parameters for the longitudinal momentum reconstruction are shown in Fig.~\ref{fig:plres}: the mean of the longitudinal momentum distribution, the root mean square, and the standard deviation. All three algorithms perform similarly, with the Recursive Annealing Fit slightly outperforming the others. The root mean square and standard deviation are lowest in the range of $0.2 - 0.5 \,$GeV/c. As stated before, low momentum particles are difficult to reconstruct with accuracy and precision. At higher momenta, the particle trajectories become straighter, introducing a larger uncertainty to the momentum reconstruction.

\subsection{DPM samples}

\begin{figure*}
  \centering
  \subfloat[]{\label{subfig:pleff_dpm}  \includegraphics[width=0.49\linewidth]{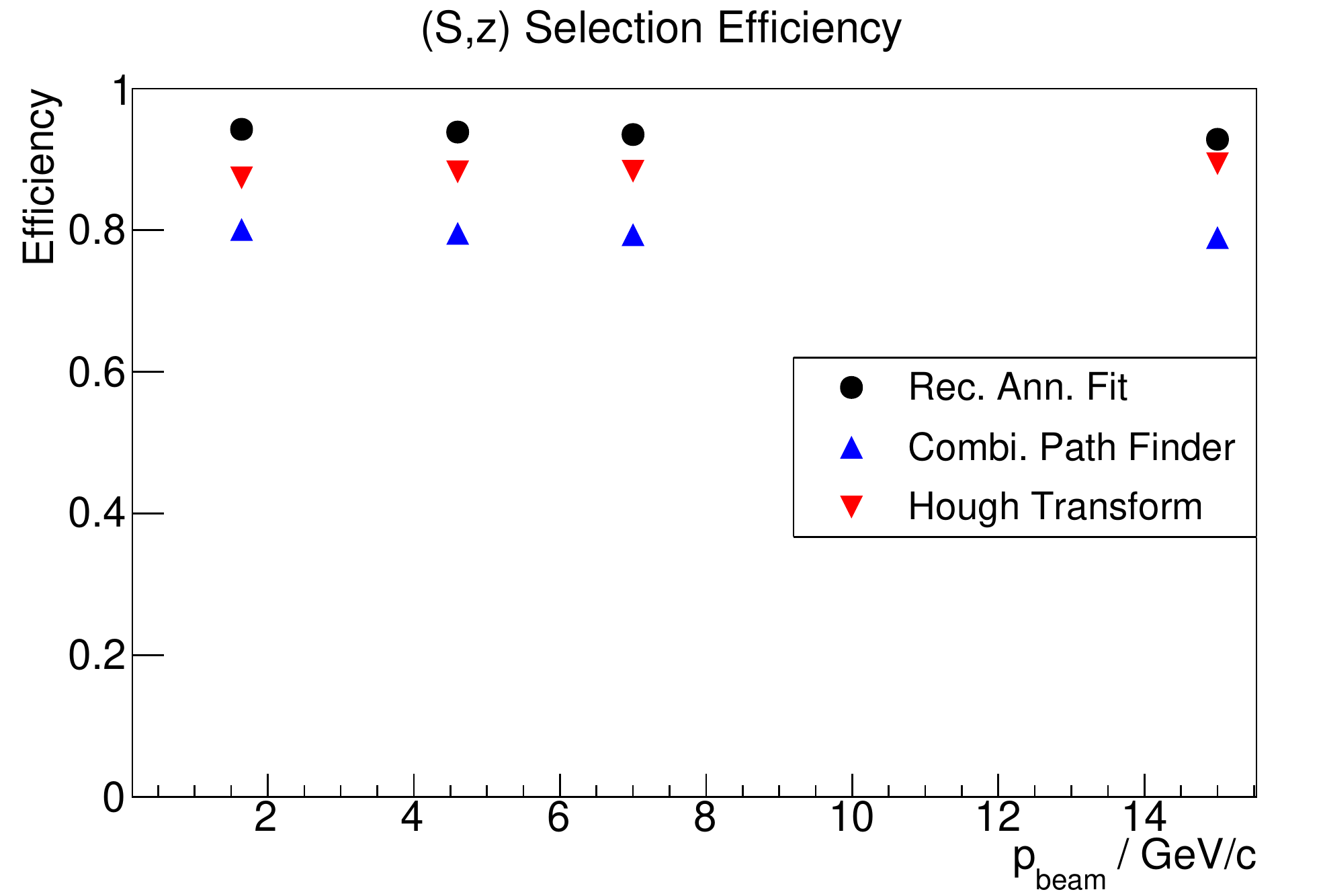}}
  \subfloat[]{\label{subfig:plpur_dpm}  \includegraphics[width=0.49\linewidth]{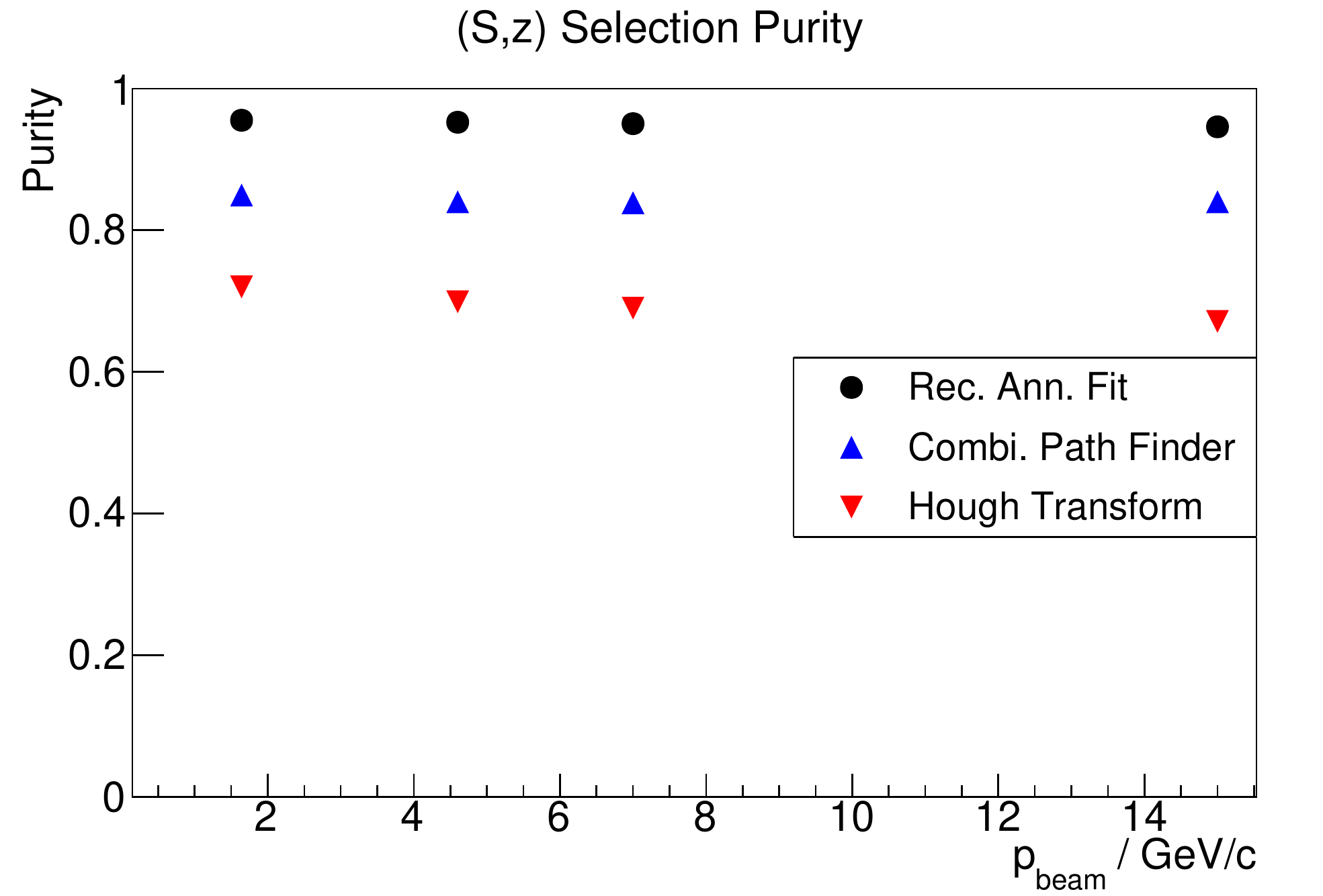}} \\
  \subfloat[]{\label{subfig:deltazMean_dpm}  \includegraphics[width=0.49\linewidth]{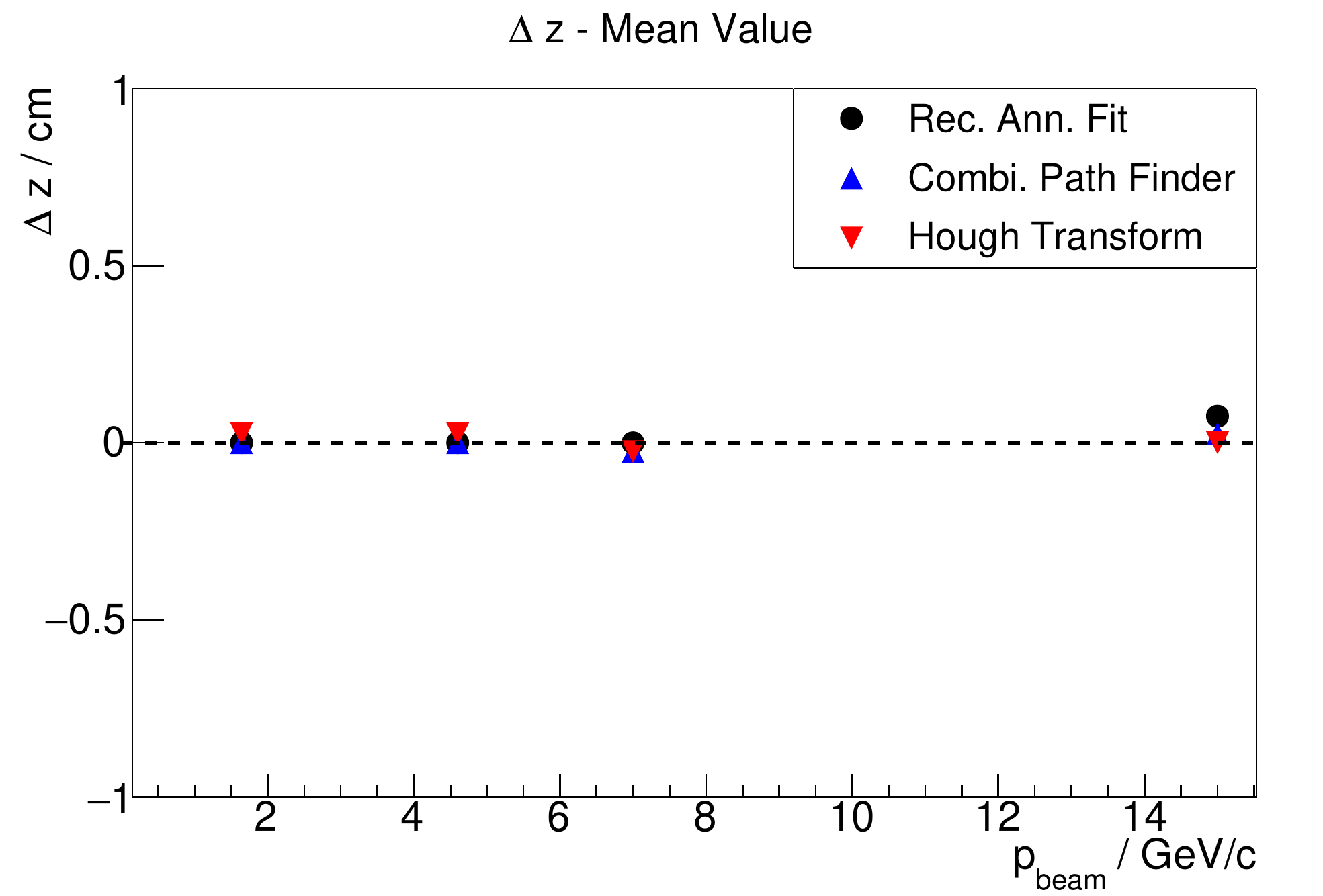}}
  \subfloat[]{\label{subfig:deltazStd_dpm}  \includegraphics[width=0.49\linewidth]{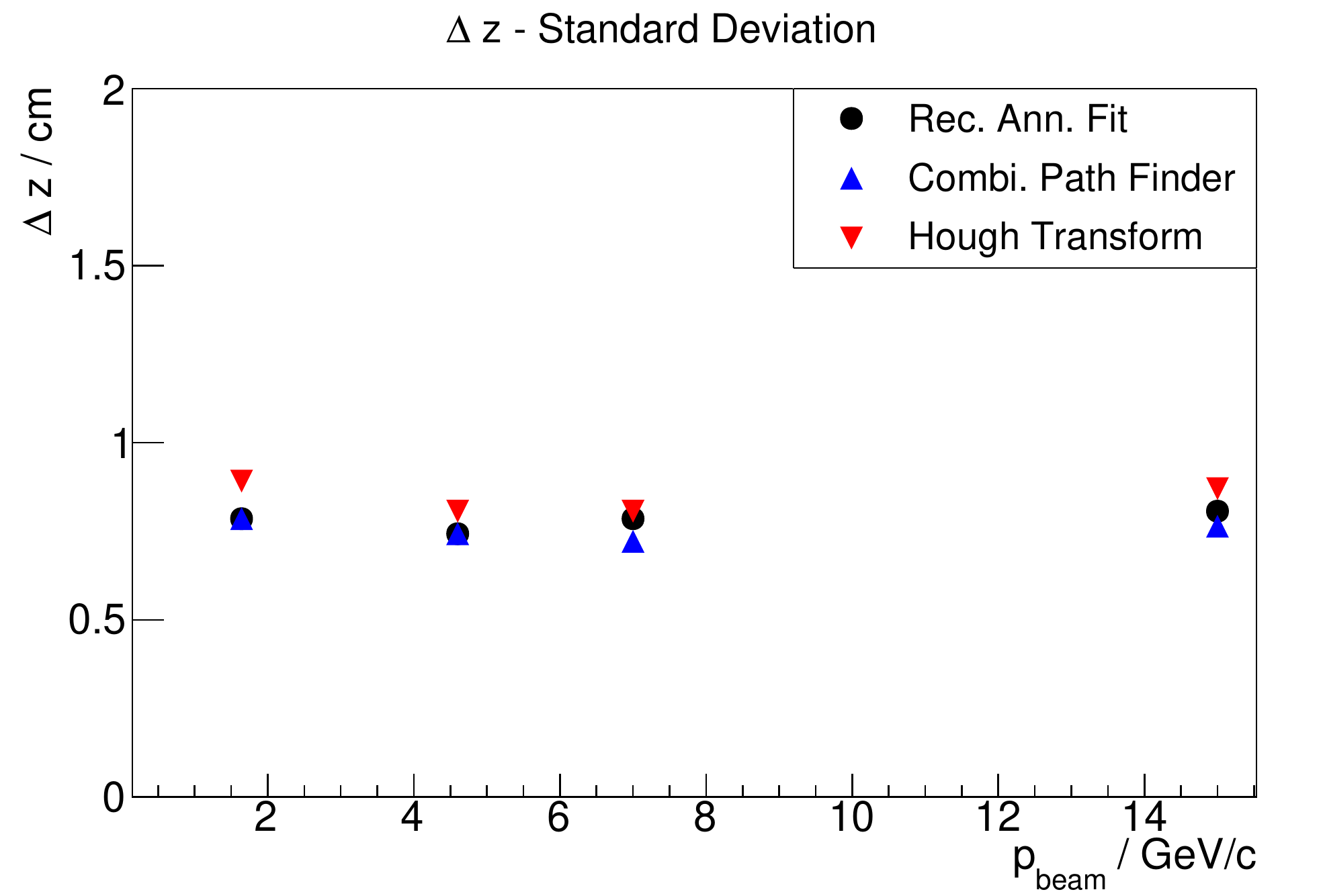}} \\
  \subfloat[]{\label{subfig:plresMean_dpm}  \includegraphics[width=0.49\linewidth]{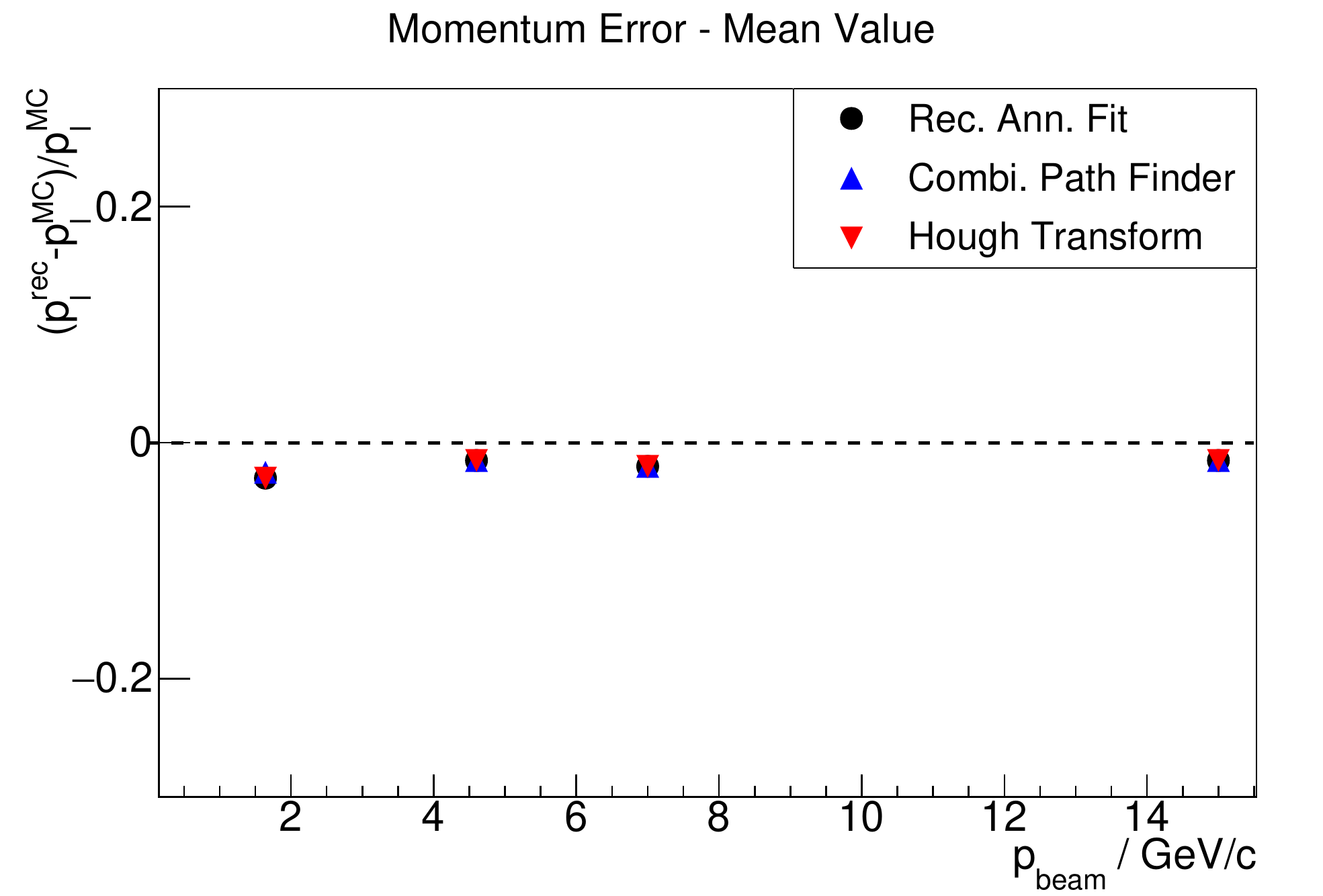}}
  \subfloat[]{\label{subfig:plresStd_dpm}  \includegraphics[width=0.49\linewidth]{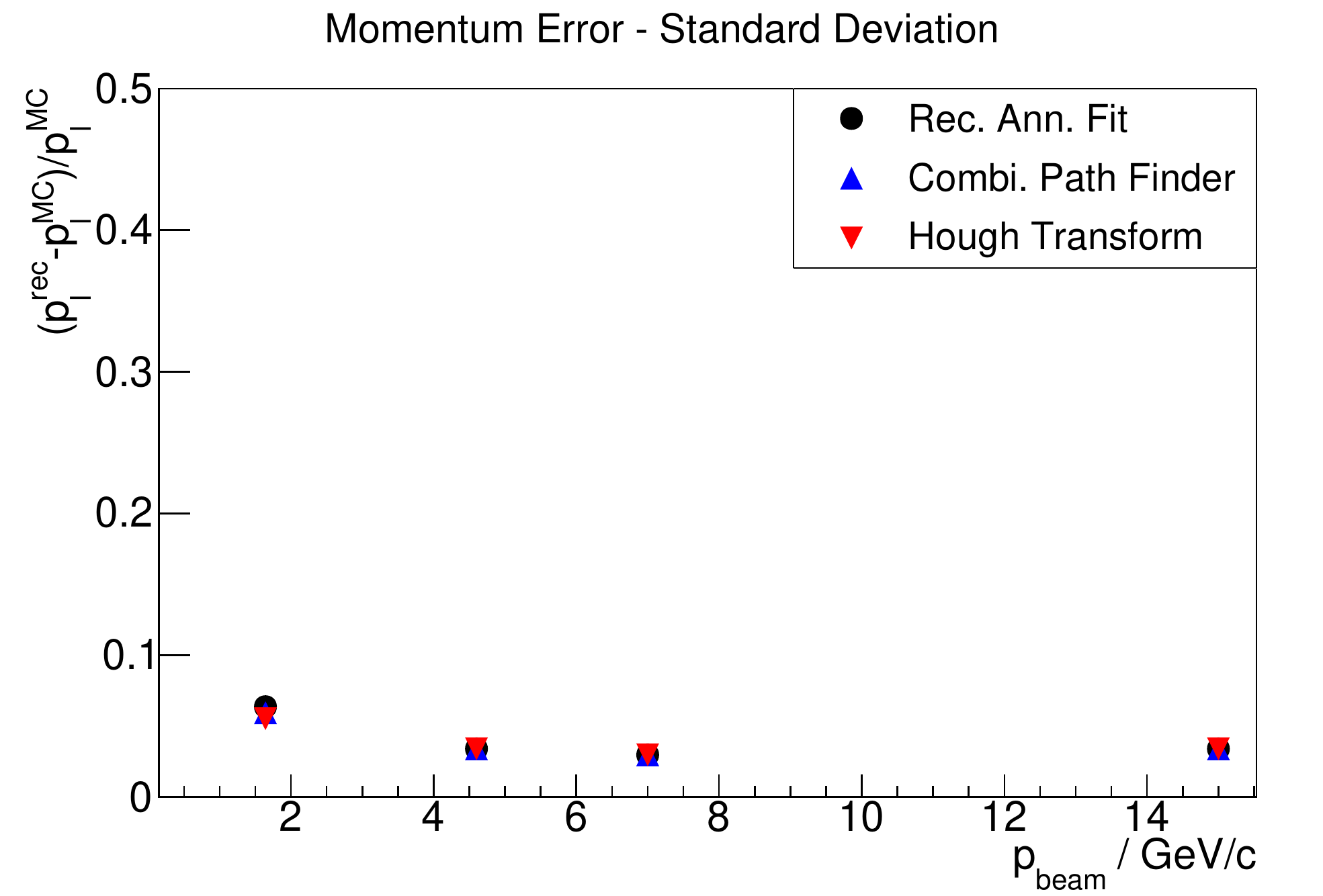}}
  \caption{Quality observables for the Recursive Annealing fit (black dots), Combinatorial Path Finder (blue upwards triangles), and Hough transformation (red downwards triangles) applied on several data samples using a DPM generator at several simulated momenta of the antiproton beam. Shown are the efficiency \ref{sub@subfig:pleff_dpm}, the purity \ref{sub@subfig:plpur_dpm}, the distribution mean for the $z$-position reconstruction \ref{sub@subfig:deltazMean_dpm} and the corresponding standard deviation \ref{sub@subfig:deltazStd_dpm}, as well as the distribution for reconstructed longitudinal momentum $p_l$ \ref{sub@subfig:plresMean_dpm} and the corresponding standard deviation \ref{sub@subfig:plresStd_dpm}.}
  \label{fig:dpm}
\end{figure*}

In addition to clean muons, the \textit{PzFinder} was tested on samples generated using a DPM generator. The corresponding quality observables are given in fig.~\ref{fig:dpm}. As before, the Recursive Annealing Fit outperforms the other methods in terms efficiency and purity, yielding values of 95\% and higher. The quality of the spatial and momentum reconstruction is comparable to the clean muon case.

\subsection{Run-time performance}

\begin{figure}[ht]
    \centering
    \includegraphics[width=\linewidth]{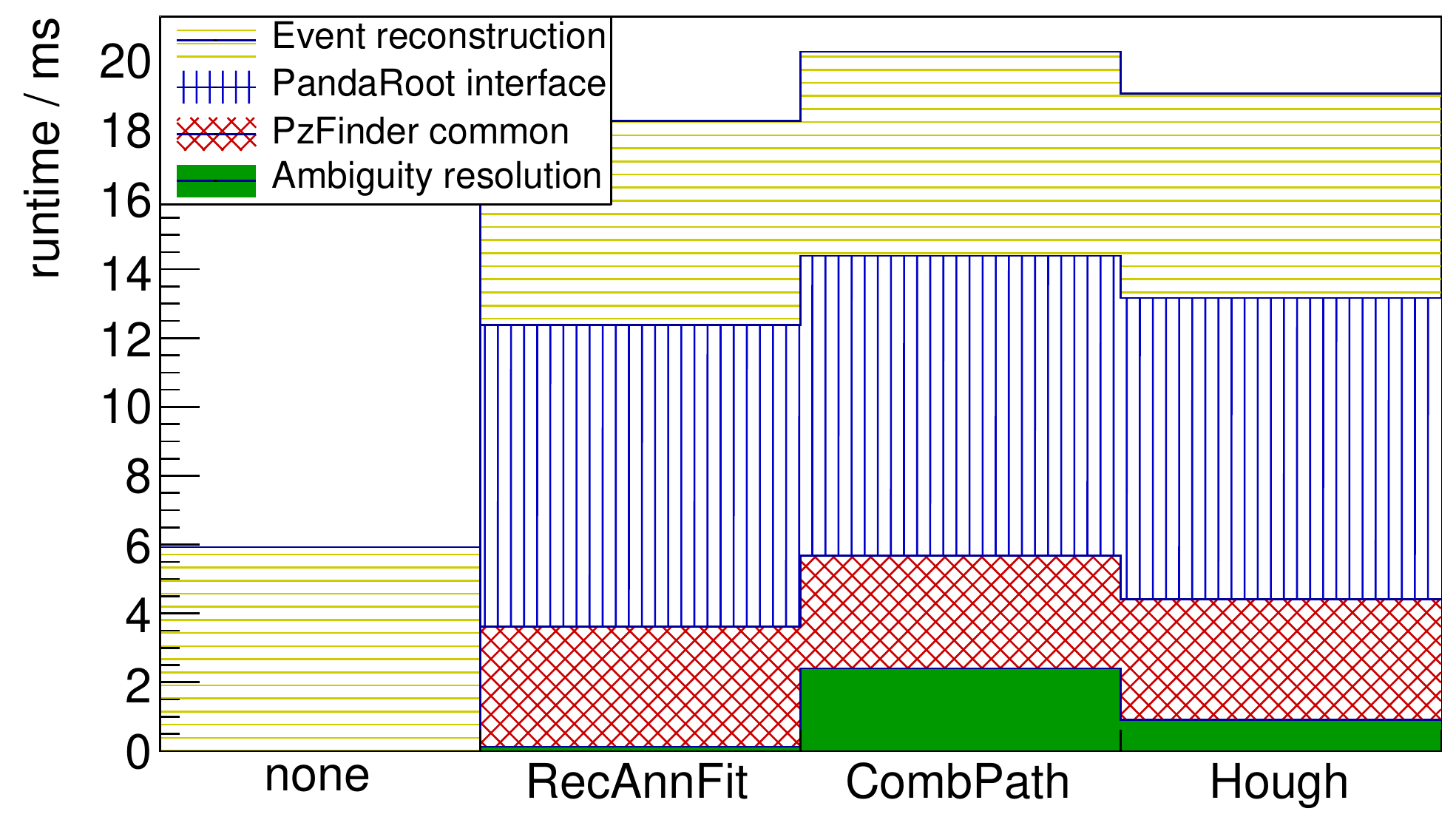}
    \caption{Run-time performance numbers in milliseconds. From left to right: no PzFinder, Recursive Annealing Fit, Combinatorial Path Finder, Hough Transformation. Add a better description here. Different parts of of the reconstruction code were measured: The preceding event reconstruction (yellow, horizontally striped), the interface between the \textit{PzFinder} and \textit{PandaRoot} (blue, vertically striped), the common part of the \textit{PzFinder} (red, checkered), and the algorithm used for resolving the left/right ambiguity (green, solid).}
    \label{fig:runtimes}
\end{figure}

Efficiency, purity, and resolution observables take a dominant role when assessing the applicability of our algorithm in an \textit{offline processing} environment, \textit{i.e.} after the experiment was carried out and the data has been taken. During \textit{online processing}, though, the reconstruction algorithms have to keep pace with the event and data rate of the ongoing experiment.

Fig.~\ref{fig:runtimes} shows timing measurements of four different cases, the event reconstruction without the \textit{PzFinder}, and then with the \textit{PzFinder} for each of its three different algorithms. The run-time is broken down into different parts of the code: The event reconstruction before the invocation of the \textit{PzFinder} (yellow, horizontally striped), the interface between the \textit{PzFinder} and \textit{PandaRoot} which consists mainly of exchanging data between the two (blue, vertically striped), the bulk of the \textit{PzFinder} code (red, checkered), which includes for example the isochrone alignment procedure and writing of output, but not the algorithm for resolving the isochrone ambiguity, which is displayed separately (green, solid). The measurements were repeated 5 times with the smallest run-time taken as the result. The machine used was an Intel Core i7-4770 CPU at 3.40 GHz.

At present, the interface between the \textit{PzFinder} and \textit{PandaRoot} adds the largest contribution to the event reconstruction time. With the goal in mind to keep the tool generically usable by multiple track finders in its prototype form, a lot of input and output data is being copied via a singleton run-time task manager. In the future, this can be mitigated or even circumvented, for example, by passing references and offering a more direct integration between the \textit{PzFinder} and other track finders. The \textit{PzFinder} sans ambiguity resolution itself adds about 60\% to the execution time. Since this is the common part of the tool, future optimisation will benefit all configurations. Among the three algorithms for ambiguity resolution, the Combinatorial Path Finder adds the largest additional run-time, close to the common part itself. The Hough transformation is about $2.5 \times$ faster. However, the Recursive Annealing Fit outperforms the others by an order of magnitude, resulting in an almost negligible computational footprint.

\section{Summary}

We have implemented and investigated three different algorithms for the reconstruction of the longitudinal momentum in the Straw Tube Tracker of the PANDA experiment. Among these three algorithms, the Recursive Annealing Fit offered the best qualitative results, yielding efficiencies and purities of 95\% and higher for particle momenta above 0.2 GeV/c. Even without adaptive fitting, the reconstructed spatial and momentum resolution comes close to the design goals set for the detector. When compared to the other two methods, the computational simplicity results in faster execution times, which adds to its appeal.

The \textit{PzFinder} is constructed in such a general way that it can be added or integrated into other track finding algorithm, that only reconstruct transversal track components, in a straightforward way. The procedure should also be applicable to other detectors of a similar geometry, for example the forward tracking stations of the PANDA detector.

The results justify further investigation as to whether the Recursive Annealing Fit should be applied to other parts of the track reconstruction chain. The extrapolation of a locally reconstructed track to another subdetector would be a possible use case.


%
%



\end{document}